\documentstyle[12pt]{article}

\def\hybrid{\topmargin 0pt      \oddsidemargin 0pt
        \headheight 0pt \headsep 0pt
        \voffset=-0.5cm
        \textwidth 6.5in        
        \textheight 9in         
        \marginparwidth 0.0in
        \parskip 5pt plus 1pt   \jot = 1.5ex}
\catcode`\@=11
\def\marginnote#1{}

\newcount\hour
\newcount\minute
\newtoks\amorpm
\hour=\time\divide\hour by60
\minute=\time{\multiply\hour by60 \global\advance\minute by-\hour}
\edef\standardtime{{\ifnum\hour<12 \global\amorpm={am}%
        \else\global\amorpm={pm}\advance\hour by-12 \fi
        \ifnum\hour=0 \hour=12 \fi
        \number\hour:\ifnum\minute<10 0\fi\number\minute\the\amorpm}}
\edef\militarytime{\number\hour:\ifnum\minute<10 0\fi\number\minute}

\def\draftlabel#1{{\@bsphack\if@filesw {\let\thepage\relax
   \xdef\@gtempa{\write\@auxout{\string
      \newlabel{#1}{{\@currentlabel}{\thepage}}}}}\@gtempa
   \if@nobreak \ifvmode\nobreak\fi\fi\fi\@esphack}
        \gdef\@eqnlabel{#1}}
\def\@eqnlabel{}
\def\@vacuum{}
\def\draftmarginnote#1{\marginpar{\raggedright\scriptsize\tt#1}}

\def\draftlabel#1{{\@bsphack\if@filesw {\let\thepage\relax
   \xdef\@gtempa{\write\@auxout{\string
      \newlabel{#1}{{\@currentlabel}{\thepage}}}}}\@gtempa
   \if@nobreak \ifvmode\nobreak\fi\fi\fi\@esphack}
        \gdef\@eqnlabel{#1}}
\def\@eqnlabel{}
\def\@vacuum{}
\def\draftmarginnote#1{\marginpar{\raggedright\scriptsize\tt#1}}

\def\draft{\oddsidemargin -.5truein
        \def\@oddfoot{\sl preliminary draft \hfil
        \rm\thepage\hfil\sl\today\quad\militarytime}
        \let\@evenfoot\@oddfoot \overfullrule 3pt
        \let\label=\draftlabel
        \let\marginnote=\draftmarginnote
   \def\@eqnnum{(\theequation)\rlap{\kern\marginparsep\tt\@eqnlabel}%
\global\let\@eqnlabel\@vacuum}  }

\def\numberbysection{\@addtoreset{equation}{section}
        \def\theequation{\thesection.\arabic{equation}}}

\def\underline#1{\relax\ifmmode\@@underline#1\else
        $\@@underline{\hbox{#1}}$\relax\fi}

\def\titlepage{\@restonecolfalse\if@twocolumn\@restonecoltrue\onecolumn
     \else \newpage \fi \thispagestyle{empty}\c@page\z@
        \def\thefootnote{\fnsymbol{footnote}} }

\def\endtitlepage{\if@restonecol\twocolumn \else  \fi
        \def\thefootnote{\arabic{footnote}}
        \setcounter{footnote}{0}}  
\relax

\numberbysection
\hybrid

\def\beq{\begin{equation}}
\def\eeq{\end{equation}}
\def\p{\partial}
\def\G{\Gamma}
\def\g{\gamma}
\def\s{\sigma}

\def\L{{\cal L}}

\def\a{\alpha}
\def\b{\beta}
\def\l{\lambda}

\def\A{{\cal A}}
\def\B{{\cal B}}
\def\AD{\A^D_{\s}}
\def\V{{\cal V}}
\def\D{{\cal D}}
\def\F{{\cal F}}
\def\GG{{\cal G}}
\def\h{{\cal H}}
\def\L{{\cal L}}
\def\K{{\cal K}}
\def\E{{\cal T}}
\def\LD{\L^D_{\g,\a}}
\def\M{{\cal M}}
\def\N{{\cal N}}
\def\O{{\cal O}}
\def\P{{\cal P}}
\def\SP{{\cal S}}
\def\ND{\N^D_{\g,\a}}

\def\NDD{\N^{D'}_{\g(x),\a(x)}}
\def\dim{{\rm dim}}
\def\res{{\rm res}}
\def\F{{\cal F}}

\def\wt{\widetilde}
\def\wh{\widehat}
\def\pp{{p}}
\def \matrix #1 {\left(\begin{array}{cc} #1 \end{array}\right)}
\newtheorem{th}{Theorem}[section]

\newtheorem{cor}{Corollary}[section]
\newtheorem{lem}{Lemma}[section]

\begin{document}

\begin{titlepage}
\title{Vector bundles and Lax equations on algebraic curves}

\author{I.Krichever \thanks{Columbia University, New York, USA and
Landau Institute for Theoretical Physics and ITEP, Moscow, Russia; e-mail:
krichev@math.columbia.edu. Research is supported in part by National Science
Foundation under the grant DMS-98-02577 and by CRDF Award RP1-2102}}

\date{August 12, 2001}

\maketitle

\begin{abstract}
The Hamiltonian theory of zero-curvature equations with spectral parameter
on an arbitrary compact Riemann surface is constructed. It is shown
that the equations can be seen as commuting flows of an infinite-dimensional
field generalization of the Hitchin system. The field analog of the elliptic
Calogero-Moser system is proposed.
An explicit parameterization of Hitchin system based on the Tyurin parameters
for stable holomorphic vector bundles on algebraic curves is obtained.

\end{abstract}

\vfill

\end{titlepage}
\newpage

\section{Introduction}
The main goal of this paper is to construct a Hamiltonian theory of
zero curvature equations on an algebraic curve introduced in \cite{kn1},
and identify them as infinite-dimensional field analogs of the Hitchin
system \cite{hit}.

The zero curvature equation
\beq\label{curv0}
\p_t L-\p_x M+[L,M]=0,
\eeq
where $L(x,t,\l)$ and $M(x,t,\l)$ are {\it rational} matrix functions of
a {\it spectral} parameter $\l$
\beq
L=u_0(x,t)+\sum_{i,s} u_{is}(x,t)(\l-\l_i)^{-s},\ \
M=v_0(x,t)+\sum_{j,k} v_{jk}(x,t)(\l-\mu_j)^{-k},
\label{M0}
\eeq
of degree $n$ and $m$, respectively,
was proposed in \cite{zakh} as one the most general type of
representation for integrable systems.
Equation (\ref{curv0}), which has to be valid identically in $\l$, is
equivalent to a system of $(n+m+1)$ matrix
equations for the unknown functions $u_0,v_0,u_{is},v_{jk}$.
The number of the equations is less than the number of
unknown functions. That is due to a gauge symmetry of (\ref{curv0}).
If $g(x,t)$ is an arbitrary matrix function then the transformation
\beq\label{gauge}
L\longmapsto g_xg^{-1}+gLg^{-1},\ \ \ M\longmapsto g_tg^{-1}+gMg^{-1}
\eeq
maps solutions of (\ref{curv0}) into solutions of the same equations.
The gauge transformation can be used to normalize $L$ and $M$. For
example, in the gauge $u_0=v_0=0$ the numbers of equations and unknown
functions are equal. Hence, equation (\ref{curv0}) is well-defined.

The Riemann-Roch theorem shows that the naive direct generalization of the
zero curvature equation for matrix functions that are meromorphic
on an algebraic curve of genus $g>0$ leads to an over-determined system of
equations. Indeed, the dimension of $(r\times r)$
matrix functions with fixed degree $d$ divisor of poles in general position
equals $r^2(d-g+1)$. If divisors of $L$ and $M$ have degrees $n$ and $m$,
then the commutator $[L,M]$ is of degree $n+m$. Therefore, the number of
equations $r^2(n+m-g+1)$ is bigger that the number $r^2(n+m-2g+1)$
of unknown functions modulo gauge equivalence.

There are two ways to overcome the difficulty in
defining the zero curvature equations on algebraic curves. The first one
is based on a choice of special ansatz for $L$ and $M$. On this way a
few integrable systems were found with Lax matrices that are elliptic
functions of the spectral parameter. The second possibility, based on
a theory of {\it high rank} solutions of the KP equation \cite{kn2},
was discovered in \cite{kn1}. It was shown that if in addition to fixed poles
the matrix functions $L$ and $M$ have moving $rg$ poles with special
dependence on $x$ and $t$, then equation (\ref{curv0}) is a well-defined
system on the space of singular parts of $L$ and $M$ at fixed poles.
Recently, an algebraic construction of the zero curvature equations on
an algebraic curve was proposed in \cite{fren}.

If matrix functions $L$ and $M$ do not depend on $x$, then (\ref{curv0})
reduces to the Lax equation
\beq\label{LL}
\p_t L=[M,L] \, .
\eeq
A theory of the Lax equations on an algebraic curve, was briefly outlined
in \cite{kn1}. In the next section for each effective degree $N>g$
divisor $D$ on a smooth genus $g$ algebraic curve $\G$ we introduce a
space $\L^D$ of the Lax matrices, and define a hierarchy of commuting
flows on it. The spaces of the Lax matrices associated to equivalent
divisors are isomorphic. If $D=\K$ is the divisor of zeros of a holomorphic
differential, then the space $\L^{\K}$ is identified with an open set of the
cotangent bundle $T^*(\wh\M)$ of the moduli space of semistable holomorphic
vector bundles on $\G$,
i.e. with an open set of the phase space of th Hitchin system.
The commuting hierarchy of the Lax equations on $\L^{\K}$ are
commuting flows of the Hitchin system.

The conventional approach to a theory of the Hitchin system is based on
a representation of $T^*(\wh\M)$ as the Hamiltonian reduction of free
infinite-dimensional system modulo infinite-dimensional gauge group.
In the finite-gap or algebro-geometric theory of soliton equations
involutivity of the integrals of motion does not come for granted, as in the case
of the Hamitonian reduction. Instead, the commutativity of the hierarchy
of the Lax equations is a starting point. It implies
involutivity of the integrals, whenever the equations are Hamiltonian.

The Lax matrices provide an explicit parameterization
of the Hitchin system based on {\it Tyurin parameters}
for framed stable holomorphic bundles on an algebraic curve \cite{tyur}.
Let $V$ be a stable, rank $r$, and degree
$rg$ holomorphic vector bundle on $\G$. Then the dimension of the space of its
holomorphic sections is $r=\dim \ H^0(\G,V)$. Let $\s_1,\ldots,\s_r$
be a basis of this space. The vectors $\s_i(\g)$ are linear independent
at the fiber of $V$ over a generic point $\g\in \G$, and are linearly dependent
\beq\label{a}
\sum_{i=1}^{r} \a_{s}^i \s_i(\g_s)=0
\eeq
at zeros $\g_s$ of the corresponding section of the determinant bundle
associated to $V$. For a generic $V$ these zeros are simple, i.e.
the number of distinct points $\g_s$ is equal to $rg=\deg \ V$, and
the vectors $\a_s=(\a_s^i)$ of linear dependence (\ref{a})
are uniquely defined up to a multiplication.
A change of the basis $\wt \s_i=\sum_j g_{ij}\s_j$ corresponds to the linear
transformation of the vectors $\a_s$, $\wt \a_s=g^T\a_s$.
Hence, an open set $\M\subset \wh \M$ of the moduli space of vector bundles is
parameterized by points of the factor-space
\beq\label{mod}
\M=\M_0/SL_r,\ \ \M_0\subset S^{rg}\left(\G\times CP^{r-1}\right) \ ,
\eeq
where $SL_r$ acts diagonally on the symmetric power of $CP^{r-1}$.
In \cite{kn1,kr2} the parameters $(\g_s,\a_s)$ were called Tyurin parameters.
Recently, the Tyurin parameterization of the Hitchin system  for $r=2$
was found \cite{rub}.

In section 3 we show that the standard scheme to solve conventional Lax
equations using the concept of the Baker-Akhiezer function is evenly
applicable to the case of Lax equations on algebraic curves.
We would like to emphasize that solution of the Lax equations
via the spectral transform of the phase space to algebraic-geometric data
does not use a Hamiltonian description of the system. Moreover, {\it a'priori}
it's not clear, why the Lax equations are Hamiltonian. In Section 4
we clarify this problem using the approach to the Hamiltonian theory of
soliton equations proposed in \cite{kp1,kp2,kr4}.
It turns out that for $D=\K$ the universal two-form which is expressed in
terms of the Lax matrix and its eigenvectors coincides with
canonical symplectic structure on the cotangent bundle $T^*(\M)$.
If the divisor $D_{\K}=D-\K$ is effective, then the form is non-degenerate
on symplectic leaves defined by a choice of the orbits of the adjoint
action of $SL_r$ on the singular parts of $L\in \L^D$ at the punctures
$P_m\in D_{\K}$.

In section 5 for each degree $N>g$ divisor $D$ on $\G$ a commuting
hierarchy of zero curvature equations is defined.
The infinite-dimensional phase space $\A^D$ of the hierarchy can
be seen as a space of connections $\p_x-L(x,q)$ along loops in $\M_0$.
We would like to emphasize that $\A^D$ does depend on the divisor $D$ and not
simply on its equivalence class, as in the case of the Lax equations.
If $D_{\K}$ is effective, then the equations of the hierarchy
are Hamiltonian after restriction on symplectic leaves.

The Riemann surface of the Bloch solutions of the equation
\beq\label{i1}
(\p_x-L(x,q))\psi(x,q)=0, \ \ x\in S^1, \ \ q\in \G
\eeq
is an analog of the spectral curves in the $x$-independent
case. Algebro-geometric solutions of the hierarchy are constructed in
the last section. Note, that they can be constructed in all the cases
independently of whether the equations Hamiltonian or not.

It is instructive to present two examples of the zero curvature equations.
The first one is a field analog of the elliptic Calogero-Moser system. The
elliptic CM system is a system of $r$ particles with
coordinates $q_i$ on an elliptic curve with the Hamiltonian
\beq\label{CM}
H={1\over 2}\left(\sum_{i}p_i^2+\sum_{i\neq j} \wp(q_i-q_j)\right),
\eeq
where $\wp(q)$ is the Weierstrass function.
In \cite{gn} the elliptic CM system was identified with
a particular case of the Hitchin system on an elliptic curve with a puncture.
In section 5 we show that the zero curvature equation on an elliptic curve
with a puncture is equivalent to the Hamiltonian system which can be seen
as the field analog of the elliptic CM system.
For $r=2$ this system is equiavalent to the system on a space of periodic
functions $p(x),q(x)$ with canonical Poisson brackets
\beq\label{br}
\{p(x),q(y)\}=\delta(x-y).
\eeq
The Hamiltonian is
\beq\label{Ham}
H=\int \left(p^2\left(1-q_x^2\right)-
{q_{xx}^2\over 2(1-q_x^2)}+2(1-3q_x^2)\wp(2q)\right)dx.
\eeq
The second example is the Krichever-Novikov equation \cite{kn2}
\beq\label{KN}
q_t={1\over 4}q_{xxx}+{3\over 8q_x}(1-q_{xx}^2)-{1\over 2}Q(q)q_x^2,
\eeq
where
\beq\label{KN1}
Q(q)=\p_q\Phi+\Phi^2,\ \ \Phi=\Phi(q,y)=\zeta(q-y)+\zeta(q+y)-\zeta(2q).
\eeq
Note, that $Q(q)$ does not depend on $y$.
Each solution $q=q(x,t)$ of (\ref{KN}) defines a rank 2, genus 1 solution of
the KP equation  by the formula
\beq\label{KN2}
8u(x,y,t)=\left(q_{xx}^2-1\right)q_x^{-2}-2q_{xxx}q_x^{-1} +8q_{xx}\Phi+
4q_x^2
\left(\p_q \Phi-\Phi\right).
\eeq
Equation (\ref{KN1}) has zero curvature representation on the elliptic curve
with puncture with $r=2$. The difference between the two examples is in the
choice of orbits at the puncture. In the first example the orbit is that of
the diagonal matrix ${\rm diag}(1,-1)$, while the second example corresponds
to the orbit of the Jordan cell.

\section{The Lax equations}

We define first the space of Lax matrices associated with a generic
effective divisor $D$ on $\G$, and a point  $(\g,\a)=\{\g_s,\a_s\}$ of
the symmetric product $X=S^{rg}\left(\G\times CP^{r-1}\right)$.
Throughout the paper it is assumed that the points $\g_s\in \G$ are
distinct, $\g_s\neq \g_k$.

Let $\F_{\g,\a}$ be the space of meromorphic vector functions $f$ on $\G$,
that are holomorphic except at the points $\g_s$, at which they have a
simple pole of the form
\beq\label{F}
f(z)={\l_s\a_s\over z-z(\g_s)}+O(1),\ \ \l_s\in C
\eeq
The Riemann-Roch theorem implies that
\beq\label{c1}
\dim\  \F_{\g,\a}\geq r(rg-g+1)-rg(r-1)=r.
\eeq
The first term in (\ref{c1}) is dimension of the space of meromorphic
vector-functions with simple poles at $\g_s$. The second term is the
number of equations equivalent to the constraint that poles of $f$ are
proportional to the vectors $\a_s$.

The space $\F_s$ of meromorphic functions in the neighborhood of $\g_s$
that have simple pole at $\g_s$ of the form (\ref{F}) is the space of
local sections of the vector bundle $V_{\g,\a}$ corresponding to $(\g,\a)$
under the inverse to the Tyurin map described in terms of Hecke modification
of the trivial bundle. The space of global holomorphic sections of
$V_{\g,\a}$ is just the space $\F_{\g,\a}$.
Let $\M_0'$ be an open set of the parameters
$(\g,\a)$ such that $\dim \ \F_{\g,\a}=r$.

Let $D=\sum_i m_i P_i$ be an effective divisor on $\G$ that
does not intersect with $\g$.
Then we define a space $\ND$ of meromorphic matrix functions
$M=M(q),\ q\in \G,$ such that:

$1^0.$ $M$ is holomorphic except at the points $\g_s$, where it has at most
simple poles, and at the  points $P_i$ of $D$, where it has poles of degree
not greater than $m_i$;

$2^0$. the coefficient $M_{s0}$ of the Laurent expansion
of $M$ at $\g_s$
\beq\label{Ms}
M={M_{s0}\over z-z_s}+M_{s1}+M_{s2}(z-z_s)+O((z-z_s)^2),\ \ z_s=z(\g_s),
\eeq
is a rank 1  matrix of the form
\beq \label{Ms0}
M_{s0}=\mu_s\a_s^T\ \longleftrightarrow M_{s0}^{ij}=\mu_s^i\a_s^j,
\eeq
where $\mu_s$ is a vector.
The constraint (\ref{Ms0}) does not depend on a choice of local
coordinate $z$ in the neighborhood of $\g_s$.

If $(\g,\a)\in \M_0'$, then the constraints (\ref{Ms0}) are linear
independent and
\beq\label{dim}
\dim\  \ND=r^2(N+rg-g+1)-r^2g(r-1)=r^2(N+1)\ ,\  N=\deg D.
\eeq
Central to all our further constructions is a map
\beq\label{map}
\D:\ND\longmapsto T_{\g,\a}\left(\M_0'\right)
\eeq
from $\ND$ to the tangent space to $\M_0'$ at the point $(\g,\a)$.
The tangent vector $\p_m=\D(M)$ is defined by derivatives of the coordinates
\begin{eqnarray}
\p_m z_s&=&-{\rm tr}\  M_{s0}=-\a_s^T\mu_s, \ \ z_s=z(\g_s),\label{t1}\\
\p_m \a_s^T&=&-\a_s^T M_{s1}+\kappa_s \a_s^T\ ,  \label{t2}
\end{eqnarray}
where $\kappa_s$ is a scalar. The tangent space to $CP^{r-1}$ at
a point represented by the vector $\a_s$ is
a space of $r$-dimensional vectors $v$ modulo equivalence
$v'=v+\kappa_s\a_s$. Therefore, the right hand side of (\ref{t2}) is a
well-defined tangent vector to $CP^{r-1}$.

Simple dimension counting shows that on an open set of $\M_0'$ the
linear map $\D$ is an injection for $N<g-1$,
and is an isomorphism for $N=g-1$.
Let us define the space $\LD$ of the Lax matrices
as the kernel of $\D$. In other words: a matrix function
$L(q)\in \ND$ is a Lax matrix if

(i) the singular term of the expansion
\beq\label{Ls}
L={L_{s0}\over z-z_s}+L_{s1}+L_{s2}(z-z_s)+O((z-z_s)^2),\ \ L_{s0}=\b_s\a_s^T,
\ \ z_s=z(\g_s),
\eeq
is traceless
\beq \label{Ls0}
\a_s^T\b_s={\rm tr}\ L_{s0}=0;
\eeq

(ii) $\a_s^T$ is a left eigenvector of the matrix $L_{s1}$
\beq\label{Ls1}
\a_s^TL_{s1}=\a_s^T\kappa_s.
\eeq
For a non-special degree $N\geq g$ divisor $D$ and a generic set of the
parameters $(\g,\a)$, the space $\LD$ is of dimension
\beq\label{dim1}
\dim\  \LD=r^2(N+1)-rg-rg(r-1)=r^2(N-g+1)\ .
\eeq
A key characterization of constraints(\ref{Ls}-\ref{Ls1}) is as follows.
\begin{lem}
A meromorphic matrix-function $L$ in the neighborhood $U$ of $\g_s$
with a pole at $\g_s$ satisfies the constraints (\ref{Ls0}) and (\ref{Ls1})
if and only if it has the form
\beq\label{lgauge}
L=\Phi_s(z)\wh L_s(z)\Phi_s^{-1}(z),
\eeq
where $\wh L_s$ and $\Phi_s$ are holomorphic in $U$, and
$\det \Phi_s$ has at most simple zero at $\g_s$.
\end{lem}
{\it Proof.}
Let  $g_s$ be a constant non-degenerate matrix such that
\beq \label{G1}
\a_s^T  g_s=e_1^T, \ e_1^T=(1,0,0,\ldots,0).
\eeq
If $L$ satisfies (\ref{Ls}, \ref{Ls0}),
then, the coefficient $L_{s0}'$ of the Laurent expansion at $\g_s$ of the
gauge equivalent Lax matrix
\beq\label{L's}
L_s'=g_s^{-1}Lg_s={L'_{s0}\over z-z_s}+L'_{s1}+O(z-z_s),\ \ z_s=z(\g_s),
\eeq
equals $fe_1^T$, where $f=g_s^{-1}\b_s$. Therefore, it has non-zero entries
at the first column, only.
\beq\label{L's1}
(L_{s0}')^{i,j}=0,\ \ j=2,\ldots,r.
\eeq
Further, the vector $e_1^T$ is a left eigenvector for
$L_{s1}'$ corresponding to the eigenvalue $\kappa_s$.
Hence, the first row of $L_{s1}'$ equals
\beq\label{L's2}
(L_{s1}')^{11}=\kappa_s,\ \ (L_{s1}')^{1j}=0,\ j=2,\ldots,r.
\eeq
From (\ref{L's1},\ref{L's2}) it follows that the matrix
$\wh L_s=f_s^{-1}L_s'f_s,$
where $f_{s}$ is the diagonal matrix
\beq\label{G}
f_{s}(z)={\rm diag}\{(z-z_s),1,1,\ldots,1\},
\eeq
is regular at $\g_s$. Hence, the Lax matrix $L$ has the
form (\ref{lgauge}), where
\beq\label{G100}
\Phi_s=g_sf_s(z)\, .
\eeq
Conversely suppose $L$ has the form (\ref{lgauge}), and
let $\a_s$ be the unique (up to multiplication) vector such that
$\a_s^T\Phi_s(z_s)=0$.
Then the Laurent expansion of $L$ at $\g_s$ has the form (\ref{Ls}).
The trace of $L$ is holomorphic, which implies  (\ref{Ls0}). Using
the equality $\a_s^T\Phi_s(z_s)=0$ we obtain that $\a_s^T L$ is
holomorphic at $\g_s$ and its evaluation at this point is proportional
to $\a_s^T$. This implies (\ref{Ls1}) and the Lemma is proved.

Let $[D]$ be the equivalence class of a degree $N>g$ divisor $D$. Then
for any set $(\g,\a)$ there is a divisor $D'$ equivalent to $D$ that
does not intersect with $\g$.
Constraints (\ref{Ls0}) and (\ref{Ls1}) are invariant under
the transformation $L\to hL$, where $h$ is a function holomorphic
in the neighborhood of $\g_s$. Therefore, the spaces $\LD$ and
$\L_{\g,\a}^{D'}$ of Lax matrices corresponding to equivalent divisors
$D$ and $D'$ are isomorphic. They can be regarded as charts of a total
space $\L^{[D]}$, the Lax matrices corresponding to $[D]$.

Let us consider in greater detail the case $D=\K$, where $\K$ is
the zero divisor of a holomorphic differential $dz$. Then $Ldz$, where
$L\in \L_{\g,\a}^{\K}$, is a matrix valued one-form
that is holomorphic everywhere except at the points $\g_s$.
The constraints (\ref{Ls0}, \ref{Ls1}) imply that the space $\F_s$ of local
sections of $V_{\g,\a}$ is invariant under the adjoint action of $L$,
\beq\label{action}
f\in \F_s\longmapsto L^T(z)f(z)\in F_s.
\eeq
Therefore, the gauge equivalence class of the matrix valued differential $Ldz$
can be seen as a global section of the bundle
$End(V_{\g,\a})\otimes \Omega^{1,0}(\G)$.
It is basic in the Hitchin system theory, that the space of such sections,
called Higgs fields, is identified with the cotangent bundle
$T^*(\wh \M)$.

It is instructive to establish directly the equivalence
\beq\label{t*}
\L^{\K}/SL_r=T^*(\M),
\eeq
using the map (\ref{map}).
The formula
\beq\label{pa}
\langle L, M\rangle=-\sum_s \res_{\g_s}{\rm Tr}\left(LM\right)dz
\eeq
defines a natural pairing between $\L_{\g,\a}^{\K}$ and $\ND$.
For a generic degree $(g-1)$ divisor $D$ the map (\ref{map}) is an isomorphism.
Therefore each tangent vector $w=(\dot z_s, \dot \a_s)$ to $\M_0'$ at
the point $(z_s=z(\g_s), \a_s)$ can be represented in the form $\D(M)$. From
(\ref{t1}, \ref{t2}) it follows that (\ref{pa}) actually defines a pairing
between  $\L_{\g,\a}^{\K}$ and the tangent space $T_{\g,\a}(\M_0)$
\beq\label{pa1}
\langle L, w\rangle=\sum_s(\kappa_s\dot z_s+\dot \a_s^T \b_s).
\eeq
This formula shows that the vector $\b_s$ and the eigenvalue
$\kappa_s$ in (\ref{Ls0}, \ref{Ls1}) can be regarded as coordinates
of a cotangent vector to $\M_0'$. Note, that $\kappa_s$ under the change of $dz$
to another holomorphic differential $dz_1$ get transformed to
$\kappa'_s=\kappa_sdz/dz_1$. Therefore, the pair $(\g_s,\kappa_s)$ can be
seen as a point of the cotangent bundle $T^*(\G)$ to the curve $\G$.

The pairing (\ref{pa1}) descents to pairing of $\L^{\K}/SL_r$ with tangent
vectors to $\M$. Indeed, tangent vectors to $\M$ at a point represented by
gauge equivalence class of $\a$ are identified with vectors $\dot \a_s$
modulo transformation $\dot \a_s^T\to \dot \a_s^T+ \a_s^TW$, where $W$
is a matrix. Under this transformation the right hand side
of (\ref{pa1}) does not changes due to the equation
\beq\label{res}
\sum_{s=1}^{rg} \b_s\a_s^T=\sum_s \res_{\g_s} Ldz=0,
\eeq
which is valid, because $Ldz$ is holomorphic except at $\g$.

The induced pairing of $\L^{\K}/SL_r$ with $T(\M)$ is non-degenerate.
Indeed, if $w=\D(M)$, then (\ref{pa}) implies that
\beq\label{pa2}
\langle L, w\rangle=\langle L, M\rangle=
\sum_i \res_{P_i}{\rm Tr}\left(LM\right)dz.
\eeq
Therefore, if (\ref{pa1}) is degenerate then there is a nontrivial
$L$ which has zero of order $m_i$ at all the points $P_i$ of $D$.
That is impossible because $D$ is a generic  degree $(g-1)$ divisor.

Our next goal is to introduce an explicit parameterization of $\L^{\K}$.
Recall, that we always assume $(\g,\a)\in \M_0'$.
\begin{lem} The map
\beq\label{100}
L\in \L^{\K}\longmapsto \{\a_s,\b_s,\g_s,\kappa_s\}\, ,
\eeq
where pairs of orthogonal vectors $(\a_s^T\b_s)=0$ are considered
modulo gauge transformations
\beq\label{101}
\a_s\to \l_s \a_s,\ \ \b_s\to \l_s^{-1}\b_s\, ,
\eeq
and satisfy equation (\ref{res}), is one-to-one correspondence.
\end{lem}
{\it Proof.} Suppose that images of $L$ and $L'$ under
(\ref{100}) coincide, then $(L-L')dz$ is a holomorphic matrix valued
differential $\varphi$ such that
\beq\label{102}
\a_s^T\varphi(\g_s)=0.
\eeq
Let $\F_{\g,\a}^P$ be the space of meromorphic
vector functions with poles at $\g_s$ of the form (\ref{F}) and
with simple pole at a point $P\in \G$. By the definition of $\M_0'$, the
constraints (\ref{F}) are linearly independent.
Therefore, $\F_{\g,\a}^P$ has dimension $2r$, and the vectors of singular
part of $f\in \F_{\g,\a}^P$ at $P$ span the whole space ${\bf C}^r$. From (\ref{102}) it follows
that if $f\in \F_{\g,\a}^P$, then the differential $f^T\varphi$ has no poles
at $\g_s$. As the sum of all the residues of a meromorphic differential equals
zero, then the $f^T \varphi$ is regular at $P$. That implies $\varphi (P)=0$.
The point $P$ is arbitrary, therefore (\ref{100}) is an injection.

The map (\ref{100}) is linear on fibers over $(\g,\a)$. Therefore, in order
to complete a proof of the lemma, it is enough to show that dimension
of $\L^{\K}_{\g,\a}$ is greater than or equal to the dimension  $d$ of
the corresponding data $(\b_s,\kappa_s)$. The vectors $\b_s$ are orthogonal
to $\a_s$.  Therefore, $d$ equals $r^2g$ minus the rank of the system
of equations (\ref{res}).

Let us show, that if $(\g,\a)\in \M_0'$ then the vectors $\a_s$ span
${\bf C}^r$.
Suppose that they  span an $l$-dimensional subspace,
then by a gauge transformation we can reduce the problem to the case
when the vectors $\a_s$ have the $(r-l)$ vanishing coordinates,
$\a_s^i=0, \ \ i>l$. The Riemann-Roch theorem then implies, that
dimension of the corresponding space $\F_{\g,\a}$ is not less than
$l(rg-g+1)-rg(l-1)+(r-l)=(r-l)g+r$.

If the rank of $\a_s^i$ is $r$, then equations (\ref{res}) are linearly
independent by themselves, but one of them is already satisfied due to
the orthogonality condition for $\b_s$, which implies
${\rm Tr} \ (\b_s\a_s^T)=0$. Therefore the dimension of the fiber of data
(\ref{100}) over $(\g,\a)\in \M_0'$ equals $r^2(g-1)+1$.

On the other hand, for $L\in \L_{\g,\a}^{\K}$
among constraints (\ref{Ls0}) there are at most
$(rg-1)$ linearly independent, because a meromorphic differential can not
have a single simple pole. Hence, dimension counting as in (\ref{dim})
implies $\dim \ \L_{\g,\a}^{\K}\ge r^2(g-1)+1$ and the Lemma is proved.

\medskip
\noindent{\bf Example.}
Let $\G$ be a hyperelliptic curve defined by the equation
\beq\label{hyp}
y^2=R(x)=x^{2g+1}+\sum_{i=0}^{2g}u_ix^i.
\eeq
A set of points $\g_s$ on $\G$ is a set of pairs $(y_s,x_s)$, such that
\beq\label{hyp1}
y_s^2=R(x_s)\ .
\eeq
A meromorphic differential on $\G$ with residues $(\b_s\a_s^T)$ at $\g_s$
has the form
\beq\label{hyp2}
L{dx\over 2y}=\left(\sum_{i=0}^{g-1}L_i x^i+\sum_{s=1}^{rg}(\b_s\a_s^T) \
{y+y_s\over x-x_s}\right)\ {dx\over 2y}\ ,
\eeq
where $L_i$ is a set of arbitrary matrices. The constraints (\ref{Ls1})
are a system of linear equations defining
$L_i$:
\beq\label{hyp3}
\sum_{i=0}^{g}\a_n^TL_i x_k^i+\sum_{s\neq n}(\a_n^T\b_s)\a_s^T \
{y_n+y_s\over x_n-x_s}=\kappa_n \a_n^T,\ \ n=1,\ldots,rg.
\eeq
in terms of  data (\ref{100}). In a similar way  the
Lax matrices can be explicitly written for any algebraic curve  using
the Riemann theta-functions.

\bigskip
For $g>1$, the correspondence (\ref{100}) descends to a system of local coordinates on
$\L^{\K}/SL_r$ over an open set $\M_0$ of $\M_0'$, which we define as
follows.

As shown above, for $(\g,\a)\in \M_0'$ the matrix $\a_s^i$
is of rank $r$. We call $(\g,\a)$ a non-special set of the Tyurin parameters
if additionally they satisfy the constraint:
there is a subset of $(r+1)$ indices
$s_1,\ldots,s_{r+1}$ such that all minors of $(r+1)\times r$ matrix
$\a_{s_j}^i$ are non-degenerate.
The action of the gauge group on the space of non-special sets of the Tyurin
parameters $\M_0$ is free.

Let us define charts of coordinates on
a smooth bundle of equivalence classes of Lax matrices over $\M_0$.
Consider the open set of $\M_0$ such that the vectors
$\a_j,\  j=1,\ldots, r,$ are linearly independent and all the coefficients
of an expansion of $\a_{r+1}$ in this basis do not vanish
\beq\label{pa33}
\a_{r+1}=\sum_{s=1}^r c_j \a_j,\ \ c_j\neq 0.
\eeq
Then for each point of this open set there exists a unique matrix
$W\in GL_r$, such that $\a_j^TW$ is proportional to the basis vector $e_j$
with the coordinates $e_j^i=\delta_j^i$, and $\a_{r+1}^TW$ is proportional to
the vector $e_0=\sum_j e_j$.
Using the global gauge transformation defined by $W$
\beq\label{pa34}
B_s=W^{-1}\b_s,\ \ A_s=W^T\a_s,
\eeq
and the part of local transformations
\beq \label{local}
A_s\to \lambda_s A_s;\ B_s\to \lambda_s^{-1}B_s,
\eeq
for $s=1,\ldots,r+1,$ we obtain that on the open set of $\M_0$
each equivalence class has representation of the form
$(A_s,B_s)$ such that
\beq\label{AA}
A_i=e_i,\ i=1,\ldots,r;\ A_{r+1}=e_0.
\eeq
This representation is unique up to local transformations (\ref{local})
for $s=r+2,\ldots,rg$.

In the gauge (\ref{AA}) equation (\ref{res}) can be easily solved for
$B_1,\ldots,B_{r+1}$. Using (\ref{AA}), we get
\beq\label{pa35}
B_j^i+B_{r+1}^i=-\sum_{s=r+2}^{rg}B_s^iA_s^j\ .
\eeq
The orthogonality condition of $B_j$ to $A_j=e_j$ implies that $B_j^j=0$.
Hence,
\beq \label{pa36}
B_{r+1}^i=-\sum_{s=r+2}^{rg}B_s^iA_s^i.
\eeq
The sets of $r(g-1)+1)$ pairs of orthogonal vectors $A_s, B_s$ modulo
the transformations (\ref{local}), and points $\{\g_s,\kappa_s\}\in S^{rg}
\left(T^*(\G)\right)$ provide a parameterization of an open set of $T^*(\M)$.
Here and below $\M=\M_0/SL_r$.

In the same way, taking various subsets of $(r+1)$ indices we obtain charts
of local coordinates which cover $T^*(\M)$.
In section 4 we provide a similar explicit parameterization of $\L^D$
for divisors $D$ such that $D_{\K}=D-{\K}$ is an effective divisor.

\bigskip

Our next goal is to construct a hierarchy of commuting flows on a total space
$\L^D$ of a vector bundle over an open set of $\M_0$.
Let us identify the tangent space $T_L(\L^D)$ to $\L^D$ at the point $L$
with the space of meromorphic matrix functions spanned by derivatives
$\p_{\tau}L|_{\tau=0}$ of all one-parametric deformations
$L(q,\tau)\in \L^D$ of $L$.

\begin{lem} The commutator $[M,L]$ of matrix functions $L\in \LD$
and $M\in \N^{D'}_{\g,\a}$ is a tangent vector to $\L^D$ at
$L$ if and only if its divisor of poles outside the points
$\g_s$ is not greater than $D$.
\end{lem}
{\it Proof.} First of all, let us show that the tangent space $T_L(\L^D)$
can be identified with a space of matrix functions
$T$ on $\G$ with poles of order not greater than $m_i$ at $P_i$, and double
poles at the points $\g_s$, where they have expansion of the form
\beq\label{T}
T=\dot z_s{\b_s\a_s^T\over (z-z_s)^2}+
{\dot\b_s\a_s^T+\b_s\dot\a_s^T\over z-z_s}+
T_{s1}+O(z-z_s).
\eeq
Here $\dot z_s$ is a constant, and $\dot \a_s, \ \dot\b_s$ are vectors
that satisfy the constraint
\beq \label{T1}
\a_s^T\dot\b_s+\dot\a_s^T\b_s=0.
\eeq
The vectors $\a_s,\b_s$ are defined by $L$. In addition it is required that the following equation holds:
\beq\label{T2}
\a_s^TT_{s1}=\dot \a_s \kappa_s +\a_s \dot \kappa_s-
\dot \a_s^T L_{s1}-\dot z_s\a_s^T
L_{s2},
\eeq
where $L_{s1},L_{s2}$ and $\kappa_s$ are defined by (\ref{Ls},\ref{Ls1}), and
$\dot \kappa_s$ is a constant.

Equations (\ref{T1}) and (\ref{T2}) can be easily checked for a tangent vector
$\p_{\tau}L|_{\tau=0}$, if we identify $(\dot z_s, \dot\a_s, \dot\b_s)$ with
\beq
\dot z_s=\p_{\tau} z(\g_s(\tau))|_{\tau=0},\ \
\dot \a_s=\p_{\tau} \a_s(\tau))|_{\tau=0},\ \
\dot \b_s=\p_{\tau} \b_s(\tau))|_{\tau=0}
\eeq
and $T_{s1}$ with
\beq
T_{s1}=\left(\p_{\tau} L_{s1}-\dot z_sL_{s2}\right)|_{\tau=0}
\eeq
Direct counting of a number of the constraints shows that
the space of matrix functions that have poles of order $m_i$ at $P_i$, and
satisfy (\ref{T}-\ref{T2}) equals $r^2(N+1)$, which is the dimension of
$\L^D$. Therefore, these relations are necessary and sufficient conditions
for $T$ to be a tangent vector.

From (\ref{Ls0}, \ref{Ls1}) it follows that, if we
define $\dot z_s$ and $\dot \a_s$ with the help of formulae (\ref{t1},
\ref{t2}), then the expansion of $[M,L]$ at $\g_s$ satisfies the constraints
(\ref{T}-\ref{T2}). The Lemma is thus proved.

The Lemma directly implies, that the Lax equation $L_t=[M,L]$ is a well-defined system
on an open set of $\L^D$, whenever we can define $M=M(L)$ as a function of
$L$ that outside of the points $\g_s$ commutes with $L$ up to a meromorphic
function with poles at the points $P_i$ of order not greater than $m_i$.

Let us fix a point $P_0\in \G$ and local coordinates $w$ in the neighborhoods
of the punctures $P_0, \ P_i\in D$. Our next goal is to define gauge
invariant functions $M_a(L)$ that satisfy the conditions of Lemma 2.3. They
are parameterized by sets
\beq\label{index}
a=(P_i,n,m),\ \ {\rm where }\ n>0, m>-m_i \ {\rm are\ integers}.
\eeq
As follows from (\ref{dim}), for generic $L\in \LD$ there is a unique matrix
function $M_a(q)$ such that:

(i)  it has the form (\ref{Ms},\ref{Ms0})  at the points $\g_s$;

(ii) outside of the divisor $\g$ it has pole at the point $P_i$, only, where
the singular part at $M_a$ coincides with the singular part of $w^{-m}L^n$,
i.e.
\beq\label{in}
M_a^-=M_a(q)-w^{-m}L^n(q)=O(1)\ {\rm is \ regular \ at \ P_i},
\eeq

(iii) $M_a$ is normalized by the condition $M_a(P_0)=0$.
\begin{th} The equations
\beq\label{La}
\p_a L=[M_a,L] ,\ \ \p_a=\p/\p t_a
\eeq
define a hierarchy of commuting flows on an open set of $\L^D$, which
descents to the commuting hierarchy on an open set of $\L^D/SL_r$.
\end{th}
By definition, $M_a$ only depends on $L$, i.e. $M_a=M_a(L)$.
Equation (\ref{in}) implies that $[M_a,L]$ satisfies the conditions of
Lemma 2.3. Therefore, the right hand side of (\ref{La}) is a tangent vector
to $\L^D$ at the point $L$. Hence, (\ref{La}) is a well-defined dynamical
system on an open set of $\L^D$.

The Laurent expansion of (\ref{La}) at $\g_s$ shows that
the projection $\pi_*(\p_a)\in T(\M_0)$ of the vector $\p_a\in T(\L^D)$
equals
\beq\label{proj}
\pi_*(\p_a)=\D(M_a)\, .
\eeq
Now let us prove the second statement of the theorem.
Commutativity of flows (\ref{La}) is equivalent to the equation
\beq\label{com}
\p_{a}M_b-\p_bM_a-[M_a,M_b]=0.
\eeq
The left hand side of (\ref{com}) equals zero at $P_0$, and, as follows
from (\ref{proj}) its expansion at $\g_s$ satisfies (\ref{T}-\ref{T2}).
Therefore, it equals zero identically, if it is regular at $D$.
This easily follows from standard arguments used in  KP theory.
If indices $a$ and $b$ correspond to the same point $P_i$, i.e.
$a=(P_i,n,m), b=(P_i,n_1,m_1)$, then in the neighborhood of $P_i$ we have
\beq\label{com11}
\p_aM_b=w^{-m_1}\p_aL^{n_1}+\p_aM_b^-=w^{-m_1}[M_a,L^{n_1}]+\p_aM_b^-=
w^{-m_1}[M_a^-,L^{n_1}]+\p_aM_b^-,
\eeq
and
\beq\label{com12}
[M_a,M_b]=[w^{-m}L^n+M_a^-,w^{-m_1}L^{n_1}+M_b^-]=
w^{-m}[L^n,M_b^-]-w^{-m_1}[L^{n_1},M_a^-]+0(1)
\eeq
From (\ref{com11}, \ref{com12}) it follows that the left hand side
of (\ref{com}) is regular at $P_i$. From the definition of $M_a$, it is regular
at all the other points of $D$ as well.
In a similar way we prove (\ref{com}) for indices $a=(P_i,n,m), \
b=(P_j,n',m')$ for $P_i\neq P_j$.

Let us now define an extended hierarchy of commuting flows on generic fibers of
the evaluation map $\LD\to L(P_0)=L_0$. Note that these fibers  are invariant
with respect to (\ref{La}). Additional flows are parameterized by indices
\beq\label{index1}
a=(P_0,m;l), \ \ m>0,\ \ l=1,\ldots, r.
\eeq
Let $L_0$ be a matrix with distinct eigenvalues, and let us fix
a representation of $L_0$ in the form $\Psi_0K_0\Psi_0^{-1}$,
where $K_0$ is a diagonal matrix.
Then for each $L\in \LD$, such that $L(q)=L_0$, there
exists a unique holomorphic matrix function
$\Psi, \Psi(q)=\Psi_0$, which diagonalizes $L$ in the neighborhood of
$q$, i.e. $ L=\Psi K \Psi^{-1}$.
For each index $a$ of the form (\ref{index1})
we define $M_a$ as the unique matrix $M_a\in \N_{\g,\a}^{nP_0}$ that
in the neighborhood of $P_0$ has the form
\beq\label{ln}
M_a=w^{-m}\Psi(w)E_{l}\Psi^{-1}(w)+O(w),
\eeq
where $E_{l}$ is the diagonal matrix $E_l^{ij}=\delta^{il}\delta^{jl}$.
\begin{th} The equations
\beq\label{Lm}
\p_aL=[M_a,L] ,\ \ a=(P_0,m;l)
\eeq
defines  commuting flows on the fiber of the evaluation map $\L^D\to L_0$,
The flows (\ref{Lm}) commute with flows (\ref{La}).
\end{th}
The proof is almost identical to that of the previous Theorem.

\section{The Baker-Akhiezer functions}

In this section we show that standard procedure in the algebro-geometric
theory of soliton equations to solve conventional Lax equations
using the concept of the Baker-Akhiezer functions (\cite{kr5,kr6}) is evenly
applicable to the case of Lax equations on algebraic curves.

Let $L\in \L^D$ be a Lax matrix. The characteristic equation
\beq\label{curve}
R(k,q)\equiv\det\left(k-L(q)\right)=k^r+\sum_{j=1}^{r}r_j(q)k^{r-j}=0
\eeq
defines a {\it time-independent} algebraic curve $\wh \G$, which is
an $r$-fold branch cover of $\G$. The following statement is a direct
corollary of Lemma 2.1.
\begin{lem} The coefficients $r_j(q)$ of the characteristic equation
(\ref{curve}) are holomorphic functions on $\G$ except at the points
$P_i$ of the divisor $D$, where they have poles of order $j m_i$, respectively.
\end{lem}
For a non-special divisor $D$ the dimension of the space $S^D$ of sets of
meromorphic functions $\{r_j(Q), \
j=1,\ldots, r\}$ with the divisor of poles $jD$
equals
\beq\label{S}
\dim \ S^D={Nr(r+1)\over 2}-r(g-1).
\eeq
Note, that dimension counting in the case of the special divisor $\K$ gives
\beq\label{SS}
\dim \ S^{\K}=r^2(g-1)+1.
\eeq
Equation (\ref{curve}) defines a map $\L^D\longmapsto S^D$. The coefficients of
an expansion of $r_j$ in some  basis of $S^D$ can be seen as functions on
$\L^D$. The Lax equation implies that these functions are integrals of motion.
Usual arguments show that they are independent. These arguments are based on
solution of {\it the inverse spectral problem}, which reconstruct $L$,
modulo gauge equivalence, from a generic
set of spectral data: a smooth curve $\wh \G$ defined by $\{r_j\}\in S^D$,
and a point of the Jacobian $J(\wh \G)$, i.e. an equivalence class $[\wh \g]$
of degree $\wh g+r-1$ divisor $\wh \g$ on $\wh \G$.
Here $\wh g$ is the genus of $\wh \G$.

For a generic point of $S$ the corresponding spectral curve $\wh\G$ is smooth.
Its genus $\wh g$ can be found with the help of the Riemann-Hurwitz formula
$2\wh g-2=2r(g-1)+\deg\nu$, where $\nu$ is the  divisor on $\G$, which  is
projection of the branch points of $\wh \G$ over $\G$. The branch points
are zeros on $\wh \G$ of the function $\p_kR(k,Q)$.
This function has poles on all the sheets of $\wh \G$ over $P_i$ of
order $(r-1)m_i$. Because the numbers of poles and zeros of a meromorphic
function are equal then $\deg\nu=Nr(r-1)$ and we obtain that
\beq\label{genus}
\wh g={Nr(r-1)\over 2}+r(g-1)+1.
\eeq
Moreover, a product of $\p_k R$ on all the sheets of
$\wh \G$ is a well-defined meromorphic function on $\G$. Its divisor of
zeros coincides with $\nu$ and the divisor of poles is $r(r-1)D$. Therefore,
these divisors are equivalent, i.e.
in the Jacobian $J(\G)$ of $\G$ we have the equality
\beq\label{j}
[\nu]=r(r-1)[D]\in J(\G).
\eeq
For a generic point $Q=(q,k)$ of $\wh \G$ there is a unique
eigenvector $\psi=\psi(Q)$ of $L$
\beq\label{p}
L(q)\psi(Q)=k\psi(Q),
\eeq
normalized by the condition that a sum of its components $\psi_i$ equals 1,
\beq\label{nor}
\sum_{i=1}^r\psi_i=1.
\eeq
The coordinates of $\psi$ are rational expressions in $k$ and the entries of $L$.
Therefore, they define $\psi(Q)$ as a meromorphic vector-function on
$\wh \G$. The degree of the divisor $\wh \g$ of its poles can be found in the
usual way. Let $\Psi(q), q\in \G,$ be a matrix with columns $\psi(Q^i)$,
where $Q^i=(q,k_i(q))$
are preimages of $q$ on $\wh \G$
\beq\label{p1}
\Psi(q)=\{\psi(Q^1),\ldots,\psi(Q^r)\}.
\eeq
This matrix depends on an ordering of the roots $k_i(q)$ of (\ref{curve}),
but the function $F(q)=\det^2\Psi(q)$ is independent of this. Therefore, $F$
is a meromorphic function on $\G$.
Its divisor of poles equals $2 \pi_*( \wh \g)$, where $\pi:\wh \G\to \G$
is the projection. In general position, when the branch
points of $\wh \G$ over $\G$ are simple, the function $F$ has simple zeros at
the images of the branch points, and double zeros the points $\g_s$, because
evaluations of $\psi$ at preimages of $\g_s$ span the subspace orthogonal
to $\a_s$. Therefore, the zero divisor of $F$ is $\nu+2\g$, where $\g=
\g_1+\cdots+\g_{rg}$, and we obtain the equality for equivalence classes
of the divisors
\beq\label{j2}
2[\pi_*(\wh \g)]=[\nu]+2[\g]=2[\g]+r(r-1)D,
\eeq
which implies
\beq\label{j3}
\deg \wh \g=\deg\nu/2+rg=\wh g+r-1.
\eeq
Let $\Psi_0$ be the matrix defined by (\ref{p1}) for $q=P_0$.
Normalization (\ref{nor}) implies that $\Psi_0$ leaves the co-vector
$e_0=(1,\ldots,1)$ invariant, i.e.
\beq\label{nor1}
e_0\Psi_0=e_0.
\eeq
The spectral curve $\wh \G$ and the pole divisor $\wh \g$ are invariant
under the gauge transformation $L\to \Psi_0^{-1}L \Psi_0,
\psi\to \Psi_0^{-1}\psi$, but the matrix $\Psi_0$ gets transformed
to the identity $\Psi_0=I$. Let $F={\rm diag} (f_1,\ldots, f_r)$ be a
diagonal matrix, then the gauge transformation
\beq
L\to FLF^{-1},\ \ \psi(Q)\to f^{-1}(Q)F\psi, \ \ {\rm where} \ \
f(Q)=\sum_{i=1}^r f_i\psi_i(Q),
\eeq
which preserves the normalization (\ref{nor}) and the equality $\Psi_0=I$,
changes $\wh \g$ to an equivalent divisor $\wh \g'$ of zeros of the meromorphic
function $f(Q)$. The gauge transformation of $L$ by a permutation matrix
corresponds to a permutation of preimages $P_0^i\in \wh \G$ of $P_0\in \G$,
which was used to define $\Psi_0$.

A matrix $g$ with different eigenvalues has  representation of the form
$g=\Psi_0F$, where $\Psi_0$ satisfy (\ref{nor1}) and $F$ is a diagonal matrix.
That representation is unique up to conjugation by
a permutation matrix. Therefore, the correspondence described above
$L\to \{\wh \G,\wh \g, \Psi_0\}$ descends
to a map
\beq\label{smap}
\L^D/SL_r\longmapsto \{\wh \G, [\wh \g]\},
\eeq
which is well-defined on an open set of $\L^D/SL_r$.

According to the Riemann-Roch theorem for each smooth genus $\wh g$
algebraic curve $\wh \G$ with fixed points $q^1,\ldots, q^r$, and
for each nonspecial degree $\wh g+r-1$ effective divisor $\wh \g$ there
is a unique meromorphic function $\psi_i(Q), Q\in \wh \G$
with divisor of poles in $\wh \g$, which is normalized by the conditions
$\psi_i(q^j)=\delta_i^j$. Let $\psi(Q)$ be a meromorphic vector-function
with the coordinates $\psi_i(Q)$. Note, that it satisfies (\ref{nor}).

Let $\wh \G$ be a curve defined by equation (\ref{curve}), where $r_j$
is a generic set of meromorphic functions on $\G$ with
divisor of poles in $jD$. Then for each point
$q\in \G$ we define a matrix $\Psi(q)$ with the help of
(\ref{p1}). It depends on a choice of order of the roots $k_i(q)$ of equation
(\ref{curve}) but the matrix function
\beq\label{inv}
L(q)=\Psi(q)K(q)\Psi^{-1}(q), \ \ K(q)={\rm diag}
(k_1(q),\ldots, k_r(q)),
\eeq
is independent of the choice, and therefore,
is a meromorphic matrix function on $\G$. It has poles of degree $m_i$ at
$P_i\in D$ and is holomorphic at the points of the branch divisor $\nu$.
By reversing the arguments used for the proof of (\ref{j3}), we get that the
degree of the zero divisor $\g$ of $\det \Psi$ equals $rg$. In general
position the zeros $\g_s$ are simple. From Lemma 2.1 it follows that an
expansion of $L$ at $\g_s$ satisfies constraints
(\ref{Ls0},\ref{Ls1}), where $\a_s$ is a unique up to multiplication
vector orthogonal to the vector-columns of $\Psi(\g_s)$.
Hence, $L$ is a Lax matrix-function.

If the points $P_0^i$ used for normalization of $\psi_j$ are preimages of
$P_0\in \G$, then $L$, given by (\ref{inv}), is diagonal at $q=P_0$, and
the correspondence $\{\wh \G,\wh \g\}\to L$
descends to a map
\beq\label{smap1}
\{\wh \G,[\wh \g]\}\to \L^D/SL_r,
\eeq
which is well-defined on an open set of the Jacobain bundle over $S$,
where it is inverse to (\ref{smap}).

Now, let $L=L(q,t)$ be a solution of the Lax equations (\ref{La},\ref{Lm}).
Then the spectral curve $\wh \G$ of $L(q,t)$ is time-independent and can be
regarded as a generating form of the integrals of the Lax equations.
The divisor $\wh \g$ of poles of the eigenvector $\psi$, defined by
(\ref{p}, \ref{nor}) does depend on $t_a$.

It is now standard procedure to show that $[\wh \g]$ evolves linearly
on $J(\wh \G)$. From the Lax equation $\p_aL=[M_a,L]$ it follows that,
if $\psi$ is an eigenvector of $L$, then $(\p_a-M_a)\psi$ is also an eigenvector.
Therefore,
\beq\label{MM}
(\p_a-M_a)\psi(Q,t)=f_a(Q,t) \psi(Q,t),
\eeq
where $f_a(Q,t)$ is a scalar meromorphic function on $\wh \G$. The
vector-function
\beq\label{Ps}
\wh \psi(Q,t)=\varphi(Q,t)\psi(Q,t),\ \
\varphi(Q,t)=\exp\left(-\int_0^{t_a} f_m(Q,\tau)d\tau\right)
\eeq
satisfies the equations
\beq\label{Ps1}
L(q,t)\wh \psi(Q,t)=k\wh\psi(Q,t),\ \
\left(\p_a-M_a(q,t)\right)\wh \psi(q,t)=0.
\eeq
It turns out that the pole divisor $\wh \g(t)$ of $\psi$ under
the gauge transform (\ref{Ps}) gets transformed  to a {\it time-independent}
divisor $\wh \g=\wh \g(0)$ of poles of $\wh \psi$. All the time dependence of
$\wh \psi(Q,t)$ is encoded in the form of its essential singularities,
which it acquires at the constant poles of $f_a$.

Let $L(q,t)$ be a solution of the hierarchy
of equations (\ref{La},\ref{Lm}). Here and below we assume that
only finite number of "times" $t_a$ are not equal to zero. For brevity
we denote the variables $t_a$ corresponding to indices (\ref{index})
and (\ref{index1}) by $t_{(i,n,m)}$ and $t_{(0,m;\,l)}$, respectively.
Commutativity of the  hierarchy implies that there is a unique common
gauge transform $\wh \psi(Q,t)=\varphi (Q,t)\psi(Q,t)$ such the
$\wh \psi$ solves all the auxiliary linear equations (\ref{Ps1}).
\begin{lem} Let $\wh \psi(Q, t), \ \wh \psi(Q,0)=\psi(Q,0)$
be the common solution of equations (\ref{Ps1}). Then

$1^0$. $\wh \psi$ is a meromorphic function on $\wh \G$ except at the points
$P_i^l$ and $P_0^l$, which are preimages on $\wh \G$ of the points $P_i\in D$
and $P_0$ on $\G$, respectively. Its divisor of poles on
$\wh \G$ outside of $P_i^l,\ P_0^l$ is not greater that $\wh \g$;

$2^0$. in the neighborhood of $P_i^l$ the function $\wh \psi$ has the form
\beq\label{e1}
\wh \psi=\xi_{i,l}(w, t)
\exp\left(\sum_n t_{(i,n,m)}w^{-m}k^n\right),
\eeq
where $\xi_{i,l}(w,t)$ is a holomorphic vector-function, and
$k=k_l(q)$ is the corresponding root of equation (\ref{curve});

$3^0$. in the neighborhood of $P_0^l$ the function $\wh \psi$ has the form
\beq\label{e2}
\wh \psi=\chi_{l}(w, t)\exp\left(\sum_n t_{(0,m;\,l)}w^{-m}\right),
\eeq
where $\chi_{l}$ is a holomorphic vector-function such that evaluation
of its coordinates at $P_0^l$ equals $\chi_{l}^i(P_0^l)=\delta^{il}.$
\end{lem}
The function $\wh\psi(Q,t)$ is a particular case of the conventional
Baker-Akhiezer functions. As shown in \cite{kr6}, for any generic divisor
$\wh \g$ of degree $\wh g+r-1$ there is a unique vector function
$\wh \psi(Q,t)$ which satisfy all the properties $1^0-3^0$.
It can be written explicitly in terms of the Riemann theta-function
of the curve $\wh \G$.
\begin{th}
Let $\wh \psi(Q,t)$ be the Baker-Akhiezer vector function associated with
a non-special divisor  $\wh \g$ on $\wh \G$. Then there exist unique
matrix functions $L(q,t), M_a(q,t)$ such that equations (\ref{Ps1})
hold.
\end{th}
As a corollary we get that the Lax operator $L(q,t)\in \L^D$ constructed with
the help of $\wh \psi$ solves the whole hierarchy of the Lax equations
(\ref{La},\ref{Lm}).

\section{Hamiltonian approach}

As we have seen, the spectral transform which identifies
the space of gauge equivalent Lax matrices with a total space  of a
Jacobian bundle over the moduli space of the spectral curves does not
involve a Hamiltonian description of the Lax equations. Moreover, {\it a'priori}
it is not clear, why all the systems constructed above are Hamiltonian.
In this section we show that the general algebraic approach to the Hamiltonian
theory of the Lax equations proposed in \cite{kp1,kp2} and developed in
\cite{kr4} is evenly applicable to the Lax equations on the Riemann surfaces.

The entries of $L(q)\in \L^D$ can be regarded as functions on $\L^D$. Therefore,
$L$ by itself can be seen as matrix-valued function and its external derivative
$\delta L$ as a matrix-valued one-form on $\L^D$. The matrix $\Psi$ (\ref{p1})
with columns formed by the canonically normalized eigenvectors $\psi(Q^i)$
of $L$ can also be regarded as a matrix function on $\L^D$ defined modulo
permutation of the columns. Hence, its differential $\delta \Psi$ is a
matrix-valued one-form on $\L^D$. In the same way we consider the
differential $\delta K$ of the diagonal matrix $K$ (\ref{inv}).
Let us define a two-form $\Omega(q)$ on $\L^D$ with values in a space
of meromorphic functions on $\G$ by the formula
\beq\label{Om}
\Omega(q)={\rm Tr} \left(\Psi^{-1} \delta L\wedge \delta \Psi-
\Psi^{-1}\delta \Psi
\wedge \delta K\right).
\eeq
This form does not depend on an order of the eigenvalues of $L$,
and therefore, is well defined on $\L^D$.
Fix a holomorphic differential $dz$ on $\G$. Then the formula
\beq\label{form}
\omega=-{1\over 2}\left(\sum_{s=1}^{rg}\res_{\g_s} \Omega dz+\sum_{P_i\in D}
\res_{P_i}\Omega dz\right),
\eeq
defines a scalar-valued two-form on $\L^D$.

The equation
\beq\label{Om1}
\delta L=\Psi \delta K \Psi^{-1}+\delta \Psi K \Psi^{-1}+
\Psi K \delta \Psi^{-1}
\eeq
implies
\beq\label{Om2}
\Omega=
2\ \delta \left({\rm Tr}\ \left(K \Psi^{-1}\delta \Psi\right)\right)=
2\ \delta \left({\rm Tr}\left(\Psi^{-1}L\delta \Psi\right)\right).
\eeq
We would like to emphasize that though the last formula looks simplier
then (\ref{Om}) and directly shows that $\omega$ is a {\it closed}
two-form, the original definition is more universal. As shown in
\cite{kp1, kp2}, it provides symplectic structure for general soliton equations.

\begin{lem} The two-form $\omega$ defined by (\ref{form}) is invariant under
gauge transformations defined by matrices $g$ that preserve the co-vector
$e_0=(1,\ldots,1),\ \  e_0g=e_0.$
\end{lem}
{\it Proof.}
If $g$ preserves $e_0$, then the gauge transformation
\beq\label{gau1}
L'=g^{-1}Lg,\ \Psi'=g^{-1}\Psi
\eeq
preserves normalization (\ref{nor}) of the eigenvectors.
If $h=(\delta g) g^{-1}$, then from (\ref{Om2}) it follows
that  under (\ref{gau1}) $\Omega$
gets transformed to $\Omega'=\Omega+F$, where
\beq\label{trans}
F=-2 \ \delta \left ({\rm Tr}\left(L h\right)\right)=-2
{\rm Tr}\left(\delta L \wedge h+L\ h\wedge h\right)
\eeq
The additional term $F$ is a meromorphic function on $\G$ with poles at
the points $\g_s$ and $P_i$.
Therefore, the sum of residues at these points of the differential $Fdz$ equals
zero and the  Lemma is proved.

It is necessary to emphasize that in the generic case the form $\omega$
is not gauge invariant with respect to the whole group $SL_r$,
because it does depend on a choice of the normalization of the eigenvectors.
A change of normalization corresponds to the transformation
$\Psi'=\Psi V, L'=L$, where $V=V(Q)$ is a diagonal matrix,
which might depend on $Q$. The corresponding transformation
of $\Omega$ has the form:
\beq\label{trans1}
\Omega'=\Omega+2 \delta \left({\rm Tr}\left(K v\right)\right)=\Omega+
2{\rm Tr}\left(\delta K\wedge v\right),\ \ v=\delta V V^{-1}.
\eeq
Here we use the equation $\delta v=v\wedge v=0$ which is valid
because $v$ is diagonal.

Let $\P^D_0\subset \L^D$ be a subspace of the Lax matrices such
that restriction of $\delta k dz$ to $\P_0^D$  is a {\it holomorphic}
differential. This subspace is a leaf of foliation on $\L^D$
defined by the common level sets of the functions defined on $\L^D$ by the
formulae
\beq\label{level}
T_{i,j,l}=\res_{P^l_i}  \left((z-z(P_i))^j\ k dz\right),\ \
j=0,\ldots, (m_i-d_i),
\eeq
where $d_i$ is the order of zero $dz$ at $P_i$ (compare with the definition of
the universal configuration space in \cite{kp1}). Note, that although the
functions (\ref{level}) are multivalued, their common level sets are
leaves of a well-defined foliation on $\L^D$.

\begin{lem} The two-form $\omega$ defined by (\ref{form}) restricted
to  $\P_0^D\subset \L^D$ is gauge invariant, i.e. it descends to a form on
$\P^D=\P_0^D/SL_r$.
\end{lem}
Let $L\in \L^{\K}$ be a Lax matrix corresponding to the zero divisor $\K$
of a holomorphic differential $dz$, then $Ldz$ has poles  at the points $\g_s$, only. Therefore,
$\P_0^{\K}=\L^{\K}$.
\begin{lem}
The two-form $\omega$ on $\L^{\K}$ defined by the formula (\ref{form})
descends to a form on $\L^{\K}/SL_r$, which under the isomorphism (\ref{t*})
coincides with the canonical symplectic structure
on the cotangent bundle $T^*(\M)$.
\end{lem}
{\it Proof.} The first statement is a direct corollary of the previous
Lemma. The second one follows from the equality
\beq\label{tr3}
\res_{\g_s} \Omega dz=-2\left(\delta \kappa_s\wedge\delta z_s+\sum_{i=1}^r
\delta \b_s^i \wedge\delta \a_s^i\right),
\eeq
which can be proved as follows.
Let $L_s'$ be the matrix defined by the gauge transformation
(\ref{L's}), and let $\Omega'_s$ be the function defined by
(\ref{Om}) for $L=L'$. Then as shown above,
\beq\label{tr4}
\res_{\g_s} \Omega_s' dz=\res_{\g_s} \Omega dz+
2\res_{\g_s}
{\rm Tr}\left(\delta g_s\ g_s^{-1}\wedge \delta L-  L\  \delta g_s \ g_s^{-1}
\wedge \delta g_s \ g_s^{-1}\right)\, .
\eeq
From (\ref{Ls},\ref{Ls0}) it follows that the second term in (\ref{tr4}) equals
\beq\label{II}
II=-2 {\rm Tr}\left((\b_s\delta \a_s^T+\delta\b_s \a_s^T)\wedge
\delta g_s\ g_s^{-1}
+\b_s\alpha_s^T \delta g_s \ g_s^{-1}
\wedge \delta g_s \ g_s^{-1}\right)\, .
\eeq
Using the equality $\delta \a_s^T g_s+\a_s^T\delta g_s=0$, which follows from
(\ref{G1}), we get
\beq\label{II1}
II=2 {\rm Tr}\left(\delta\b_s\wedge \delta \a_s^T\right)=
-2 \left(\delta\a_s^T\wedge \delta \b_s\right).
\eeq
The matrix $L_s'$ under the gauge transformation $\wh L=f_s^{-1}L_s'f_s$,
where $f_s=f_s(z)$ is the diagonal matrix (\ref{G}), gets transformed to
a holomorphic matrix. Therefore,
\beq\label{III}
0=\res_{\g_s} \Omega_s' dz+
2\res_{\g_s}
{\rm Tr}\left(\delta f_s\ f_s^{-1}\wedge \delta L'_s-  L_s'\  \delta f_s \ f_s^{-1}
\wedge \delta f_s f_s^{-1}\right).
\eeq
The last term in (\ref{III}) equals zero because $f_s$ is diagonal.
From (\ref{G1}-\ref{G}) it follows that
\beq\label{III1}
\res_{\g_s} \Omega_s' dz=
-2\res_{\g_s}{\rm Tr}\left(\delta f_s\ f_s^{-1}\wedge \delta L_s'\right)=
2\delta z_s\wedge \delta \kappa_s.
\eeq
Equations (\ref{tr4}-\ref{III1}) imply (\ref{tr3}). In the
coordinates $A_s$ and $B_s$ (\ref{pa34}-\ref{pa36}) on an open set of
$T^*(\M)$ the form $\omega$ due to (\ref{AA}) equals
\beq\label{tr55}
\omega_0=\sum_{s=1}^{rg}\delta \kappa_s\wedge\delta z_s+\sum_{s=r+1}^{rg}
\delta B_s^T \wedge\delta A_s, \ g>1
\eeq
and the Lemma is proved.

Let us now consider the contribution to $\omega$ from poles of
$Ldz$ at the points $P_m$ of the divisor $D_{\K}=D-{\K}$. The residue of the
last term in (\ref{Om}) restricted to $\P_0^D$ vanishes. Therefore,
\beq\label{P1}
\omega_m=-{1\over 2}\res_{P_m}\Omega dz=\res_{P_m}{\rm Tr}\left(
L\delta \Psi\Psi^{-1}\wedge \delta \Psi\Psi^{-1}\right)dz\, .
\eeq
If $Ldz$ has a simple pole at $P_m$, then its residue $L_m$ is a point
of the orbit $\O_m$ of the adjoint action of $GL_r$, corresponding to
the fixed singular part of $k dz$, which defines the leaf $\P_0^D$.
Let $\xi$ be a matrix, which we regard as a point of the Lie algebra
$\xi\in sl_r$. The formula
\beq\label{P2}
\p_{\xi} L_m=[L_m,\xi],
\eeq
defines a tangent vector $\p_{\xi}\in T_{L_m}(\O_m)$ to the orbit at
$L_m$. The correspondence ${\xi}\to\ \p_{\xi}$ is a isomorphism between
$sl_r/sl_r(L_m)$, and $T_{L_m}(\O_m)$. Here $sl_r(L_m)$ is a subalgebra of
the matrices, that commute with $L_m$. Evaluation of the form
$\left(\delta \Psi \Psi^{-1}\right)$ at $\p_{\xi}$ is equal to
$\xi$. Hence, (\ref{P1}) restricted to $\P_0^D$ coincides with the canonical
symplectic structure on the orbit $\O_m$. Its evaluation on a pair of vectors
${\xi},\eta$ is equal to
\beq\label{P3}
\omega_m({\xi},\eta)={\rm Tr} \left(L_m\ [\xi,\eta]\right).
\eeq
If $Ldz$ has a multiple pole at $P_m$, then we define $\wt L_m$ as the
equivalence class of the singular part of $Ldz$. By definition two
matrix differentials $\wt L$ and $\wt L'$ meromorphic in the neighborhood of
$P_m$  are equivalent if $\wt L-\wt L'$ is a holomorphic differential.
Let $\GG_-$ be a group of the invertible holomorphic matrix functions in the
neighborhood of $P_m$. The transformation
$\wt L\to g\wt L g^{-1},\ \ g\in \GG_-$ defines a
representation of $\GG_-$ on the finite-dimensional space of singular
parts of meromorphic differentials. Let $\wt{\O}_m$ be an orbit of this
representation.

If $\h_-$ is the Lie algebra of $\GG_-$, then the equivalence class of the
right hand side of (\ref{P2}) for $\xi\in \h_-$ depends only on the
equivalence class of $\wt L_m$. Therefore, (\ref{P2}) defines an isomorphism
between the tangent space to $\wt{\O}_m$ at
$\wt L_m$ and $\h_-/\h_-(\wt L_m)$, where $\h_-(\wt L_m)$ is the subalgebra
of holomorphic matrix functions ${\xi}$ such that $[L_m,{\xi}]$ is
holomorphic at $P_m$. The formula
\beq\label{P4}
\omega_m=\res_{P_m} {\rm Tr} \left(\wt L_m\ [{\xi},{\eta}]\right)
\eeq
defines a symplectic structure on $\wt {\O}_m$.

\begin{lem} If $D_{\K}=D-{\K}>0$ is an effective divisor, then the map
\beq\label{B1}
L \longmapsto \{z_s,\kappa_s, \a_s,\b_s, \wt L_m,\},
\eeq
is a bijective correspondence between points of the bundle $\L^D$ over $\M_0$
and sets of the data (\ref{B1}) subject to the constraints
$(\a_s^T\b_s)=0$, and
\beq\label{B2}
\sum_{s=1}^{rg} \b_s\a_s^T+\sum_{P_m\in D'}\res_{P_m} \ \wt L_m=0,
\eeq
modulo gauge transformations (\ref{101}).
\end{lem}
If we fix a gauge on a open set of $\L^D$ by (\ref{AA}), then the
reconstruction formulae for $B_1,\ldots, B_{r+1}$ become
\beq \label{pa36a}
B_{r+1}^i=-\sum_{s=r+2}^{rg}B_s^iA_s^i-\sum_m \res_{P_m}\, \wt L_m^{ii},
\eeq
and
\beq\label{pa35a}
B_j^i=-B_{r+1}^i-\sum_{s=r+2}^{rg}B_s^iA_s^j-\sum_m \res_{P_m}\, \wt L_m^{ij}
\, .
\eeq
If $g>1$, then for $D_{\K}>0$ the data
$\{z_s,\kappa_s, A_s, B_s, \wt L_m\in \wt \O_m \}$ provide explicit
coordinates on an open set of $\P^D$.
\begin{th} Let $D$ be a divisor such that $D_{\K}\geq 0$, where $\K$ is
the zero divisor of a holomorphic differential $dz$. Then the form $\omega$
defined by (\ref{form}), restricted to $\P^D_0$ descends to non-degenerate
closed two-form on $\P^D$:
\beq\label{B3}
\omega=\omega_0+\sum_{P_m\in D_{\K}} \omega_m,
\eeq
where $\omega_0$ and $\omega_m$ are given by is by (\ref{tr55}), and
(\ref{P4}), respectively.
\end{th}
The representation of the form $\omega$ in terms of the Lax operator and
its eigenvectors provides a straightforward and universal way to show
that the Lax equations are Hamiltonian, and to construct the action-angle
variables.

By definition a vector field $\p_t$ on a symplectic
manifold is Hamiltonian, if the contraction $i_{\p_t}\omega(X)=
\omega(\p_t,X)$ of the symplectic form
is an exact one-form $dH(X)$. The function $H$ is the Hamiltonian
corresponding to the vector field $\p_t$.

\begin{th}
Let $\p_a$ be the vector fields corresponding to the Lax equations (\ref{La},
\ref{Lm}). Then the contraction of $\omega$ defined by (\ref{form})
restricted to $\P^D$ equals
\beq
i_{\p_{a}}\omega=\delta H_{a},
\eeq
where
\begin{eqnarray}
H_{a}&=&-{1\over n+1}\ \res_{P_i}{\rm Tr}\left(w^{-m} L^{n+1}\right) dz ,  \
a=(P_i,n,m),
\label{H11} \\
H_{a}&=&-\res_{P_0}\ \left(w^{-m}k_l\right) dz, \ \ \ \ \
a=(P_0,m;\, l),\ \  \label{H10}
\end{eqnarray}
Here $k_l=k_l(q)$ is the $l$-th eigenvalue of $L$ in the
neighborhood of the puncture $P_0$.
\end{th}
\noindent{\it Proof.}
The Lax equation $\p_a L=[M_a,L]$, $\p_a k=0$, and equation (\ref{MM})
\beq
\p_a\Psi=M_a\Psi+\Psi F_a , \label{d}
\eeq
where $\Psi$ is the matrix of eigenvectors (\ref{p1}), and  $F_a={\rm diag}
(f_a(Q^1),\ldots,f_a(Q^r)$,  imply
\beq\label{H1}
i_{\p_a}\omega=-{1\over 2}\left(\sum_{s=1}^{rg}\res_{\g_s}\Lambda dz+
\sum_{P_i\in D}\res_{P_i}\Lambda dz\right),
\eeq
where $\Lambda=\Lambda(q)$ equals
\beq\label{H2}
\Lambda={\rm Tr}\left(\Psi^{-1}[M_a,L]\delta \Psi-
\Psi^{-1}\delta L(M_a\Psi+\Psi F_a)-
\Psi^{-1}(M_a\Psi+\Psi F_a)\delta K\right).
\eeq
Using, as before, the equality
$L\delta \Psi-\delta \Psi K=\Psi \delta K-\delta L \Psi$,
we get that
\beq\label{H3}
{\rm Tr}\left(\Psi^{-1} [M_a,L]\delta \Psi\right)\ =
{\rm Tr}\left(\Psi^{-1}M_a\Psi \delta K- M_a\delta L\right).
\eeq
Using the fact that $K$ and $F$ are diagonal, we also obtain the equation
\beq\label{H4}
{\rm Tr}\left(\Psi^{-1} \delta L
\Psi\ F_a\right)={\rm Tr}\left(\delta K\ F_a\right).
\eeq
From (\ref{H3},\ref{H4}) it follows that
\beq\label{H6}
i_{\p_a}\omega=\sum_{P_i\in D}\res_{P_i}{\rm Tr}\
\left( \delta K  \ F_a\right)dz+R_a,
\eeq
where
\beq\label{H7}
R_a=\sum_{s=1}^{rg}\res_{\g_s}{\rm Tr}\left(\delta LM_a\right) dz+
\sum_{P_i\in D}\res_{P_i}{\rm Tr}\left(\delta LM_a\right)dz.
\eeq
Note that in the first term of (\ref{H6})
a sum of residues at $\g_s$ has been dropped because $K$ and $F_a$
are holomorphic at these points.

Consider first the case of the Lax equations (\ref{La}).
The matrix $M_a$ for $a=(P_i,n,m)$ is holomorphic everywhere
except at the points $\g_s$ and $P_i$. Therefore, $R_{i,n,m}=0$.
The corresponding diagonal matrix $F_{i,n,m}$
is holomorphic at the points $P_j\in D,\ j\neq i$. From
(\ref{in}) it follows that $F_{i,n,m}$ in the neighborhood of
$P_i$ has the form
\beq\label{H8}
F_{i,n,m}=-w^{-m}K^n+O(1).
\eeq
The form $\delta Kdz$ restricted to $\P^D$ is holomorphic in the
neighborhood of $P_i$. Therefore,
\beq\label{H9}
-\res_{P_i}{\rm Tr}\left( \delta K  F_{i,n,m}\right)dz=
\res_{P_i}{\rm Tr}\left(w^{-m} K^n\delta K \right)dz=
{1\over n+1}\res_{P_i}{\rm Tr}\left(w^{-m}L^{n+1}\right)dz.
\eeq
The matrix $F_a$ corresponding to $a=(P_0,m;\, l)$ is
holomorphic at the points of $D$. Therefore, the right hand side of (\ref{H6})
reduces just to $R_a$. Because, $M_{0,m;\, l}$ is holomorphic except at
the points $\g_s$ and $P_0$, we have in this case the equation
\beq\label{H15}
R_{0,m;\, l}=-\res_{P_0}{\rm Tr}\left(\delta LM_{0,m;\, l}\right) dz,
\eeq
which, with the help of (\ref{ln}), implies (\ref{H10}).
The Theorem is therefore proved. It shows that the Lax equations restricted
to $\P^D$ are Hamiltonian whenever the restriction of $\omega$
is non-degenerate.
\begin{cor} If $D_{\K}$ is an effective divisor, then the Lax equations
(\ref{La},\ref{Lm}) restricted to $\P^D$ are Hamiltonian.
The corresponding Hamiltonians (\ref{H11},\ref{H10}) are in involution
\beq\label{HHH}
\{H_a,H_b\}=0.
\eeq
\end{cor}
The basic relation which implies all equations (\ref{HHH}) is
involutivity of all the eigenvalues of the Lax matrices
at different points of $\G$, i.e.
\beq\label{HH1}
\{k_l(q),k_{l_1}(q_1)\}=0.
\eeq
\noindent{\bf Example.} Let us consider the Lax matrices
on an elliptic curve $\G=C/\{2n\omega_1,2m\omega_2\}$ with one puncture,
which without loss of generality we put at $z=0$.
In this example we denote the parameters $\g_s$ and $\kappa_s$ by $q_s$ and
$p_s$, respectively.

In the gauge $\a_s=e_s,\ e_s^j=\delta_s^j$ the $j$-th column of
the Lax matrix $L^{ij}$ has poles only at the points $q_j$ and $z=0$.
From (\ref{Ls0}) it follows that $L^{jj}$ is regular everywhere, i.e. it
is a constant. Equation (\ref{Ls1}) implies that $L^{ji}(q_j)=0,\ i\neq j$
and $L^{jj}=p_j$. An elliptic function with two poles and one zero
fixed is uniquely defined up to a constant. It can be written in terms
of the Weierstrass $\s$-function as follows
\beq\label{W}
 L^{ij}(z)=f^{ij}{\s(z+q_i-q_j)\, \s(z-q_i) \s(q_j) \over
\s(z)\s(z-q_j)\,\s(q_i-q_j)\,\s(q_i)}, \ i\neq j;\ \ L^{ii}=p_i.
\eeq
Let $f^{ij}$ be a rank 1 matrix $f^{ij}=a^ib^j$. As it was mentioned above,
the equations $\a_i=e_i$ fix the gauge up to transformations by diagonal
matrices. We can use these transformation to make $a^i=b^i$.
The corresponding momentum is given then by the collection $(a^i)^2$ and we
fix it to the values $(a^i)^2=1.$ The matrix $L$ given by (\ref{W}) with
$f^{ij}=1$ is gauge equivalent to the Lax matrix $\wt L$ with a spectral
parameter for the elliptic Calogero-Moser system found in \cite{kr7}:
\beq\label{cm}
\wt L^{ii}=p_i,\ \ \wt L^{ij}=\Phi(q_i-q_j,z),\ i\neq j,
\eeq
where
\beq\label{cm1}
\Phi(q,z)={\s(z-q)\over \s(z)\s(q)}\ e^{\zeta(z)q}
\eeq
Note that $\wt L$ has essential singularity at $z=0$, which is due to the
gauge transformation by the diagonal matrix
$\wh \Phi={\rm diag} (\Phi(q_i,z)$, which removes poles of $L$ at the
points $q_i$.

The Hamiltonian of the elliptic CM system (\ref{CM}) is equal to
\beq\label{cm2}
H_{CM}={1\over 2}\  \res_{0} \ {\rm Tr}\left(z^{-1}L^2\right)dz.
\eeq
For the sequel, we would like to express $H_{CM}$
in terms of the first two coefficients of the Laurent expansion of
the marked branch of the eigenvalue of $L$ at $z=0$. Indeed, expansions of the eigenvalues of $L$ at
$z=0$ have the form
\beq\label{cm3}
k_1(z)=(r-1)z^{-1}+k_{11}+k_{12}z+O(z^2),\ \
k_l(z)=-z^{-1}+k_{l1}+k_{l2}z+O(z^2) , l>1.
\eeq
The equation
\beq\label{cm4}
H_1=\sum_{i=1}^rp_i={\rm Tr}\ L=\sum_{l=1}^r k_l(z).
\eeq
implies
\beq\label{cm5}
H_1=\sum_{l=1}^r k_{l1},\ \ \sum_{l=1}^r k_{l2}=0.
\eeq
From (\ref{cm3}) and (\ref{cm5}) it follows that
\beq\label{cm6}
2H_{CM}=2rk_{12}+\sum_{l=1}^r k_{l1}^2.
\eeq
Trace of $L^m$ has the only pole at $z=0$. Hence, we have the equations
\begin{eqnarray}
\res_{0}{\rm Tr}\  (L^2)&=&2\left((r-1)k_{11}
-\sum_{l=2}^r k_{l1} \right)=0,
\label{cm7}     \\
\res_{0}{\rm Tr} (L^3)&=&3\left((r-1)^2k_{12}+(r-1)k_{11}^2+
\sum_{l=2}^r \left(k_{l2}-k_{l1}^2\right) \right)=0.
\label{cm8}
\end{eqnarray}
Equations (\ref{cm7}) and (\ref{cm8}) imply
\beq
H_1=rk_{11},\ \ 2H_{CM}=r^2k_{12}+rk_{11}^2.
\eeq
Our next goal is to construct the action-angle variables for $\omega$.
\begin{th}
Let $L\in \L^D$ be a Lax matrix, and let $\wh \g_s$ be the poles of the
normalized (\ref{nor}) eigenvector $\psi$. Then
the two-form $\omega$ defined by (\ref{form}) is equal to
\beq\label{d1}
\omega=\sum_{s=1}^{\wh g+r-1} \delta k(\wh \g_s)\wedge \delta z(\wh \g_s).
\eeq
\end{th}
The meaning of the right hand side of this formula is as follows.
The spectral curve is equipped by definition
with the meromorphic function $k(Q)$. The pull back to $\wh \G$ of
the abelian integral $z(Q)=\int^{Q}dz$ on $\G$ is a multi-valued holomorphic
function on $\wh \G$.
The evaluations $k(\wh \g_s),\ z(\wh \g_s)$ at the points
$\wh \g_s$ define functions on the space $\L^D$, and the wedge product of
their external differentials is a two-form on $\L^D$. (Note, that
differential $\delta z(\wh \g_s)$ of the multi-valued function
$z(\wh g_s)$ is single-valued, because the periods of $dz$ are constants).

\noindent{\it Proof.}
The proof of the formula (\ref{d1}) is very general
and does not rely on any specific form of $L$. Let us present it briefly
following the proof of Lemma 5.1 in \cite{kr4} (more details can be found in
\cite{spin}).

Let $\g_s^j, P_i^j$ be preimages on $\wh \G$ of
the points $\g_s\in \G$ and $P_i\in D$. Then the form $\omega$ is equal to
\beq\label{df}
\omega=
-{1\over 2}\sum_{j=1}^r
\left(\sum_{s=1}^{rg}\res_{\g_s^j} \wt\Omega dz+\sum_i
\res_{P_i^j}\wt \Omega dz\right),
\eeq
where $\wt \Omega$ is a meromorphic function on $\wh \G$ defined by the
formula
\beq\label{df1}
\wt \Omega(Q)=\psi^{*}(Q) \delta L(q)\wedge \delta \psi(Q)-
\psi^{*}(Q) \delta  \psi(Q)
\wedge \delta k,\ \ Q=(k,q)\in \wh \G.
\eeq
The expression $\psi_n^*(Q)$ is the dual eigenvector, which is
the row-vector solution of the equation
\beq\label{dual}
\psi^*(Q)L(q)=k\psi^*(Q),
\eeq
normalized by the condition
\beq\label{nord}
\psi^*(Q)\psi(Q)=1.
\eeq
Note that $\psi^*(Q)$ can be identified with the only row of the matrix
$\Psi^{-1}(q)$ which is not orthogonal to the column $\psi (Q)$
of $\Psi(q)$.
That implies that $\psi^*(Q)$ as a function on the spectral curve
has poles at the points $\g_s^j$, and at the branching points of the spectral
curve. Equation (\ref{nord}) implies that it has zeroes at the poles $\wh \g_s$
of $\psi_n(Q)$. These analytical properties will be crucial in the sequel.

The differential $\wt \Omega dz$ is a meromorphic differential on the
spectral curve $\wh \G$. Therefore, the sum of its residues at the punctures
$P_{i}^j,\ \g_s^j$ is equal to the negative of the sum of the other
residues on $\wh \G$.
There are poles of two types. First of all, $\wt \Omega$ has poles at the poles
$\wh \g_s$ of $\psi$. Note that $\delta \psi$ has pole of the second order
at $\wh \g_s$. Taking into account that $\psi^*$ has zero at $\wh \g_s$
we obtain
\beq
\res\,_{\hat \g_s}\wt \Omega=(\psi^*\delta L\psi)(\wh \g_s)
\wedge \delta z(\wh \g_s)+\delta k(\wh \g_s)\wedge\delta z(\wh \g_s)=
2\delta k(\wh \g_s)\wedge\delta z(\wh \g_s). \label{65}
\eeq
The last equality follows from the standard formula for variation of the  eigenvalue of an  operator, $\psi^*\delta L\psi=\delta k$.

The second set of poles of $\wt \Omega$ is the set of branch points
$q_i$ of the cover. The pole of $\psi^*$  at $q_i$ cancels with the zero
of the differential $dz, \ dz(q_i)=0$, considered as differential on $\wh \G$.
The vector-function $\psi$ is holomorphic at $q_i$. If we take an expansion of
$\psi$ in the local coordinate $(z-z(q_i))^{1/2}$
(in general position when the branch point is simple) and consider its
variation we get that
\beq
\delta \psi=-{d\psi\over dz}\delta z(q_i)+O(1).\label{66}
\eeq
Therefore, $\delta \psi$ has simple pole at $q_i$. In the similar way
we have
\beq
\delta k=-{dk\over dz} \delta z(q_i). \label{67}
\eeq
Equalities (\ref{66}) and (\ref{67}) imply that
\beq
\res\,_{q_i}
\left(\psi^{*} \delta L\wedge \delta \psi\right)dz=
\res_{q_i}\left[ (\psi^*\delta L d\psi)\wedge
{\delta k dz\over
dk}\right]\ . \label{68}
\eeq
Due to skew-symmetry of the wedge product we we may replace $\delta L$ in
(\ref{68}) by $(\delta L-\delta k)$. Then, using the identities
$\psi^*(\delta L-\delta k)= \delta \psi^* (k-L)$  and
$(k-L)d\psi=(dL-dk)\psi$, we obtain
\beq
\res_{q_i}
\left(\psi^{*} \delta L\wedge \delta \psi\right)dz
=-\res_{q_i}(\delta \psi^*\psi)\wedge \delta k dz=
\res_{q_i}(\psi^*\delta \psi)\wedge \delta k dz.
\label{000}
\eeq
Note that the term with $dL$ does not contributes to the residue, because
$dL(q_i)=0$. The right hand side of (\ref{000}) cancels with a residue of
the second term in the sum (\ref{df1}) and the Theorem is proved.

\noindent{\it Remark}. The right hand side of (\ref{d1}) can be identified
with a particular case of universal algebraic-geometric symplectic form
proposed in \cite{kp1}. It is defined on the generalized Jacobian bundles
over a proper subspaces of the moduli spaces of Riemann surfaces with
punctures. In the case of families of hyperelliptic curves that form was
pioneered by Novikov and Veselov \cite{nv}.

Let $\phi_k$ be coordinates on the Jacobian $J(\wh \G)$ of the spectral
form. The isomorphism of the symmetric power
of the spectral curve and the Jacobian is defined by the Abel map
\beq\label{ab0}
\phi_i(\wh \g)=\sum_{s}\int^{\wh \g_s} d\omega_i,
\eeq
where $d\omega_i$ is the basis of normalized holomorphic differentials on
$\wh \G$, corresponding to a choice of a basis of $a$- and $b$-cycles on
$\wh \G$ with the canonical matrix of intersections.
Restricted to $\P^D$, the differential $\delta kdz$ is holomorphic.
Therefore, it can be represented as a sum of the basis differentials
\beq\label{ab}
\delta kdz=\sum_i\delta I_i d\omega_i\, .
\eeq
The coefficients of the sum are differentials on $\P^D$ of the functions
\beq\label{A}
I_i=\oint_{a_i}kdz\, .
\eeq
From (\ref{d1}) it follows that $\omega=\delta \a$, where
\beq\label{h70}
\a=\sum_{s=1}^{\wh g+r-1} \int^{\wh \g_s} \delta k dz=
\sum_{i=1}^{\wh g} \delta I_i\wedge\phi_i.
\eeq
\begin{cor} The form $\omega$ restricted to $\P^D$ equals
\beq\label{form3}
\omega=\sum_{i=1}^{\wh g} \delta I_i\wedge\delta\phi_i.
\eeq
\end{cor}
For the case, when $D_{\K}\geq 0$, this result was obtained first
in \cite{nek2}.

It is instructive to show that (\ref{form3}) directly implies that $\omega$
is non-degenerate for $D_{\K}\ge 0$. First of all, (\ref{form3}) implies that
the forms $\delta I_i$ are linear independent. Indeed, if they are linear
dependent at $s\in S^D$, then there is a vector $v$ tangent to $S^D$ at
$s$, such that $\delta I_i(v)=0$. Due to (\ref{ab}) we conclude
$\p_v k\equiv 0$. It is impossible for generic $s$,  because
the equation
\beq\label{H20}
\p_v k={\sum_{j=1}^{r}\p_v r_jk^{r-j}\over R_k(k,Q)}\equiv 0,
\eeq
implies, then, that  $k(Q)$ satisfies algebraic equation of degree less
than $r$, i.e. the spectral curve $\wh \G$ can not be $r$-sheeted branch
cover of $\G$.

Second argument needed in order to complete the proof is that dimension
of the space $S_{\P}^D\subset S^D$ of the spectral curves corresponding to
$\P^D$ equals $\wh g$. The number of conditions that singular
parts of eigenvalues of $L$ at the points $P_m\in D_{\K}$ are constant
along $\P^D$ equals $(r\deg D_{\K})$ minus $1$, due to the relation
\beq\label{H50}
\sum_{P_m\in D_{\K}}  \res_{P_m} ({\rm Tr}\  L)dz=0,
\eeq
which is valid, because the singular parts of $L$ at $\g_s$ are traceless.
From (\ref{S}) we get
\beq\label{SS1}
2\ \dim \ S^D_{\P}=Nr(r-1)-2r(g-1)+2=2\ \wh g=\dim \ \P^D.
\eeq

\section{The zero-curvature equations}

The main goal of this section is to present the non-stationary analog
of the Lax equations on an algebraic curve as an infinite-dimensional
Hamiltonian system.

Let $\A^D$ be a space of $(r\times r)$ matrix function $L(x,q)=L(x+T,q)$
of the real variable $x$ such that:

$1^0.$ $L(x,q)$ is a meromorphic function of the variable $q\in \G$
with poles at $D$ and at the points $\g_s(x)$, where it has
the form (\ref{Ms}), i.e. $L(x,q)\in \N^D_{\g(x),\a(x)}$,
\beq\label{Ms1}
L(x,z)={\b_s(x)\, \a^T(x)\over z-z_s(x)}+L_{s1}(x)+O((z-z_s(x)),\ \
z_s(x)=z(\g_s(x)),
\eeq

$2^0.$ the vector $\D(L(x,q))$ defined by map (\ref{map})
is {\it tangent} to the loop $\{\g(x),\a(x)\}$, i.e.
\beq
\p_x z_s(x)=-\a_s^T(x)\,  \b_s(x),\
\p_x \a_s^T(x)=-\a_s^T(x) L_{s1}(x)+\kappa_s(x) \a_s^T(x)\ ,  \label{t21}
\eeq
where $\kappa_s(x)$ is a scalar function.

\noindent{\it Remark.} It is necessary to emphasize, that although
the loops $S^1\longmapsto \N^D/SL_r$ are lifted to matrix functions
$L'(x,q)\in \N^D,\ x\in R,$ such that
\beq\label{z00}
L'(x+T,q)=gL'(x,q)g^{-1}+\p_xg g^{-1},\ \ \g=g(x)\in GL_r,
\eeq
without loss of generality we may consider fucntions periodic in $x$,
because $L'$ with the monodromy property (\ref{z00}) is gauge equivalent
to a periodic matrix function $L$.

The space $\AD$ of the matrix functions, corresponding to a loop
$\s=\{\g(x),\a(x)\}$ in $\M_0$, is the space of sections of
finite-dimensional affine bundle over the loop, because for any two functions
$L_1,L_2\in \AD$  their difference is the Lax matrix,
$L_1-L_2\in \L^D$. Therefore, for a generic
divisor $D$ the space $\AD$ is non-trivial only if $\deg D=N\geq g$.
The functional dimension of $\AD$ is equal to $r^2(N-g+1)$, while the
functional dimension of $\A^D$ equals $r^2(N+1)$.
\begin{lem} If $D=\K$ is the zero divisor of a holomorphic
differential $dz$, then the map
\beq\label{a1}
L\in \A^{\K} \longmapsto \{\a_s(x),\b_s(x),\g_s(x),\kappa_s(x)\}
\eeq
is a bijective correspondence of $\A^{\K}$ and the space of functions
periodic in $x$ such that
\beq \label{a2}
\p_xz(\g_s(x))=-\a_s^T(x)\b_s(x),\ \ \sum_{s=1}^{rg}\b_s(x)\a_s(x)^T=0,
\eeq
modulo the gauge transformations
\beq\label{a3}
\a_s(x)\longmapsto \l_s(x)\a_s(x),\  \b_s\longmapsto \l_s^{-1}(x)\b_s(x),\
\kappa_s(x)\longmapsto \kappa_s(x)+\p_x\ln \l_s(x)
\eeq
\beq\label{a4}
\a_s(x)\longmapsto W(x)^T\a_s(x),\ \ \b_s(x)\longmapsto W^{-1}(x)\b_s(x),
\eeq
where $\l_s(x)$ is a non vanishing function periodic in $x$ and
$W(x)\in \wh{GL_r}$ a periodic non-degenerate matrix function.
\end{lem}
Note that from (\ref{t21}) it follows that locally in the neighborhood of
$\g_s(x)$ the matrix function $L(x,Q)\in \AD$ can be regarded as connection of the
bundle $\wh \V$ over $S^1\times \G$ along the loop $\{\g(x),\a(x)\}$.
Indeed, if $\F$ is a space of local sections
of this bundle, which can be identified with the space of meromorphic
vector functions $f(x,z)$ that have the form (\ref{F}) in the neighborhood of
$\g_s$, then
\beq\label{A1}
\left(\p_x+L^T(x,z)\right)f(x,z) \in \F_s.
\eeq
Another characterization of the constraints (\ref{t21}) is as follows.
\begin{lem} A meromorphic matrix-function $L$ in the neighborhood of $\g_s(x)$
with a pole at $\g_s(x)$ satisfies the constraints (\ref{t21}) if and only
if there exists a holomorphic matrix function $\Phi_s(x,z)$ with
at most a simple zero of $\det \Phi_s$ at $\g_s$ such that
$L$ is gauge equivalent
\beq\label{agauge}
L=\Phi_s \wh L \Phi_s^{-1}+\p_x\Phi_s \ \Phi_s^{-1}
\eeq
to a holomorphic matrix function $\wh L$.
\end{lem}
The tangent space to $\A^D$ is the space of functions of $x$ with values
in the tangent space to the space of Lax matrices $T(\L^D)$.
\begin{lem} Let $L\in \AD$ and $M\in \NDD$, then the commutator
$[\p_x-L, M]=M_x+[M,L]$ is a tangent vector to $\A^D$ at
$L$ if and only if its divisor of poles outside of $\g_s(x)$ is
not greater than $D$.
\end{lem}
From equations (\ref{t21}) it follows that the Laurent expansion of
the matrix function $T=M_x+[L,M]$ at the point $\g_s(x)$ has the form
(\ref{T}), where $\dot z_s$ and $\dot \a_s$ are given by
formulae (\ref{t1}, \ref{t2}). That proves that $T$ is a tangent vector
to $\L^D$.

Lemma 5.3 shows that the zero-curvature equation
\beq\label{z1}
L_t=M_x+[M,L]
\eeq
is a well-defined system, whenever we can define $M(L)$, such
that the conditions of the Lemma are satisfied. Our goal is to construct
the zero-curvature equations that are equivalent to {\it differential
equations}. That requires $M(L)$ to be expressed in terms of $L$ and its
derivatives in $x$.

It is instructive enough to consider the case when all the multiplicities
of the points $P_i\in D$ equal $m_i=1$. Let $\A^D_0$ be an open set in $\A^D$
such that the singular part of $L\in \A^D_0$ at $P_i$ has different
eigenvalues
\begin{eqnarray}
L(x,q)&=&w_i^{-1}C_i(x)\ u^{(i)}(x) C_i^{-1}(x)+O(1), \ \
w_i=w_i(q), \ \ w_i(P_i)=0, \nonumber \\
u^{(i)}&=&
{\rm diag} \left(u^{(i)}_1(x),\ldots, u^{(i)}_r(x)\right),\ \
u^{(i)}_k(x)\neq u_l^{(i)}(x), \ \ k\neq l. \label{z1a}
\end{eqnarray}
\begin{lem}
Let $L(x,w)$ be a formal Laurent series
\beq\label{z2}
L=\sum_{j=-1}^{\infty} l_{j}(x)w^j
\eeq
such that $l_{-1}(x)=C(x) u(x) C^{-1}(x)$, where $u$ is a diagonal matrix,
with distinct diagonal elements. Then there is a unique formal solution
$\Psi_0=\Psi_0(x,w)$ of the equation
\beq\label{z4}
\left(\p_x-L(x,w)\right)\Psi(x,w)=0,
\eeq
which has the form
\beq\label{z5}
\Psi_0(x,w)=
C(x)\left(\sum_{s=0}^{\infty}\xi_s(x)w^{s}\right)
e^{\int_{x_0}^x h(x',w)dx'},   \ \  h={\rm diag} (h_1,\ldots,h_r),
\eeq
normalized by the conditions
\beq\label{z6}
\xi_{0}^{ij}=\delta^{ij},\ \ \xi^{ii}_s(x)=0.
\eeq
The coefficients $\xi_s(x)$ of (\ref{z5}) and the coefficients $h_s(x)$
of the Laurent series
\beq\label{z51}
h(x,w)=\sum_{s=-1}^{\infty} h_s(x)w^s , \ \ h_{-1}=u,
\eeq
are differential polynomials of the matrix elements of $L$.
\end{lem}
Substitution of (\ref{z5}) into (\ref{z4}) gives a system of
the equations, which have the form
\beq\label{z9}
h_s-[u,\xi_{s+1}]=R(\xi_0,\ldots,\xi_{s};h_0,\ldots,h_{s-1}),\ \ \  s=-1,0,1,\ldots.
\eeq
They recursivelyly determine the off-diagonal part of $\xi_{s+1}$, and the
diagonal matrix $h_s$ as polynomial functions of matrix elements
of $l_i(x), \ i\leq s.$
\begin{cor} Let $\Psi_0$ be the formal solution (\ref{z5}) of
equation (\ref{z4}). Then for any diagonal matrix $E$ the expression
$w^{-m}\Psi_0E\Psi_0^{-1}$
does not depend on $x_0$, and is formally meromorphic, i.e. it has the form
\beq\label{z90}
w^{-m}\Psi_0E\Psi^{-1}_0=
\sum_{s=-m}^{\infty} m_s(x)w^{-s} .
\eeq
The coefficients $m_s(x)$ are differential polynomials on the matrix elements
of the coefficients $l_i(x)$.
\end{cor}
Expression (\ref{z90}) is meromorphic and does not depend on
$x_0$, because the essential singularities of the factors commute with $E$
and so cancel each other.

We are now in position to define matrices $M_a$,
\beq\label{f1}
a=\left(P_i,m;l\right), \ \ m\geq 1,\ \ l=1,\ldots,r,
\eeq
which are differential polynomials on entries of $L$, and
satisfy the conditions of Lemma 5.3.
Let $\Psi_0(x,q)=\Psi_0(x,w(q))$ be the formal solution of equation (\ref{z4})
constructed above for the expansion (\ref{z1a}) of $L\in \A_0^D$
at $P_i$. Then, we define $M_{(i,m;l)}(x,q)$ as the unique
meromorphic matrix function, which  has the form
(\ref{Ms}, \ref{Ms0}) at the points $\g_s(x)$, and is holomorphic everywhere else except
at the point $P_i$, where
\beq\label{z12}
M_{(i,m;l)}(x,q)=w^{-m}(q)\Psi_0(x,q) E_l\Psi_0^{-1}(x,q)+O(1),  \ \
E_l^{ij}=\delta^{i}_l\delta^{jl}.
\eeq
As before, we normalize $M_{(i,m;l)}$ by the condition
$M_{(i,m;l)}(x,P_0)=0$.

It is necessary to mention, that $M_a$, as a function of $L$, is defined
only locally, because it depends on a representation of the singular
part of $L$ at $P_i$ in the form (\ref{z1a}).
\begin{th}
The equations
\beq\label{z13}
\p_a L=\p_x M_a+[M_a,L] , \ \ a=(P_i,m;l)
\eeq
define a hierarchy of commuting flows on $\A^D_0$.
\end{th}
Let the coefficients of (\ref{z2}) be periodic functions of $x$.
Then, Lemma 5.4 implies that
\beq\label{z10}
\Psi_0(x+T,w)=\Psi_0(x,w)e^{p(w)},\ \
p=\int_{0}^T h(x,w)dx.
\eeq
Therefore, the columns of $\Psi_0$ are Bloch solutions of equation
(\ref{z4}), i.e. the solutions that are eigenvectors of the monodromy
operator. The diagonal elements of the matrix $\pp(w)$ are the formal
quasimomentum of the operator (\ref{z4}).

Our next goal is to show that for $D_{\K}\geq 0$ the zero curvature
equations are Hamiltonian on suitable symplectic leaves,
and identify their Hamiltonians with coefficients of the quasimomentum
matrices $\pp_i$ corresponding to the expansion (\ref{z1a}) of $L$ at the punctures $P_i$
\beq\label{z11}
\pp_{i}(w)=\sum_{s=-1}^{\infty} H_{(i,s)}w^s, \ \
H_{(i,s)}={\rm diag}\ \{H_{(i,s;l)}\},
\eeq
Let us fix a holomorphic differential $dz$ with simple zeros,
and a set of diagonal matrix functions $v^{(i)}(x)$.
Then for a divisor $D$, such $D_{\K}$ is effective, we define first a
subspace $\B^D$ of $\A^D_0$ by the constraints
\beq\label{zz1}
\p_x\left(u^{(i)}(x)-v^{(i)}(x)\right)=0,
\eeq
where $u^{(i)}$ are the matrices of eigenvalues (\ref{z1a}) of the singular
parts of $L\in \A^D_0$. Next we define a foliation of $\B^D$. The leaves
$\wh \P_0^D$ of the foliation are parameterized by sets of constant
diagonal matrices $c^{(m)}$ with distinct diagonal elements, and are
defined by the equations
\beq\label{z14}
u^{(m)}(x)-v^{(m)}(x)=c^{(m)},  \ \ {\rm if}\ \ \ dz(P_m)\neq 0,
\eeq
We would like to stress the difference between the constraints
(\ref{zz1}) and (\ref{z14}). Equations (\ref{zz1}) imply that for all
the points of the divisor $D$ the differences $(u^{(i)}(x)-v^{(i)}(x))$
are  ${\it x-independent}$ matrices. For $P_m\in D_{\K}$
we requier additionally that the difference equals to the fixed matrix.

As before, we define a two-form on $\wh\P_0^D$ by formula (\ref{form}),
where now
\beq\label{z15}
\Omega(q)={\rm Tr} \left(
\int_{x_0}^{x_0+T}\left(\Psi^{-1} \delta L\wedge \delta \Psi\right)dx
-\left(\Psi^{-1}\delta \Psi\right)(x_0)\wedge \delta \pp\right)
\eeq
and $\Psi$ is the matrix of the Bloch solutions of (\ref{z4}), i.e.
\beq\label{z16}
(\p_x-L(x,q))\Psi(x,q)=0,\ \
\Psi(x+T,q)=\Psi(x,q)e^{\pp\,(q)}.
\eeq
We would like to emphasize that this definition is a slight modification
of the formula for symplectic structure for soliton equations, proposed
in \cite{kp1}. The second term in (\ref{z15}) gives zero contribution
in the conventional theory. It is here to remove the dependence on the choice
of $x_0$ in the definition as may be seen as follows.
The monodromy property (\ref{z16}) implies
\beq\label{z17}
{\rm Tr}\left(\Psi^{-1} \delta L\wedge \delta \Psi\right)(x+T)-
{\rm Tr}\left(\Psi^{-1} \delta L\wedge \delta \Psi\right)(x)=
{\rm Tr}\left(\left(\Psi^{-1} \delta L\Psi\right)(x)
\wedge \delta \pp\right).
\eeq
Using the equations $\delta L\Psi=\delta \Psi_x-L\delta \Psi$,
we obtain
\beq\label{z18}
{\rm Tr}
\left(\Psi^{-1} \delta L\Psi\right)=
{\rm Tr}\left(\p_x\left(\Psi^{-1} \delta \Psi\right)\right).
\eeq
Hence, the form $\Omega$ does not depend on a choice of the
initial point $x=x_0$.

The same arguments as before show that $\omega$ when restricted to
$\wh \P_0^D$ does not depend on the normalization of the Bloch solutions.

\begin{th} The formula (\ref{form}) with $\Omega$ given by (\ref{z15}) defines
a closed two-form on $\wh \P_0^D$. This is gauge invariant with respect to
the affine gauge group $\wh {GL_r}$.

If $D\ge\K$, then the contraction of $\omega$ by the vector field $\p_a$
defined by (\ref{z13}) equals
\beq\label{z23}
i_{\p_{a}}\omega=\delta H_{a},
\eeq
where for $a=(P_i,m;l)$
\beq\label{z181}
H_{(i,m;l)}=-\res_{P_i} {\rm Tr} \ \left(w^{-m}E_l\ \pp\right) dz,
\eeq
and $\pp$ is the quasi-momentum matrix.
\end{th}
The proof of this theorem proceeds along identical lines to the proof of the
stationary analogs of these results presented above. First, we show that
under the gauge transformation $L'=g^{-1}Lg-g^{-1}\p_x,\ \Psi'=g^{-1}\Psi$
the form $\Omega$ gets transformed to
\beq\label{z19}
\Omega'=\Omega+{\rm Tr}\ \int_{x_0}^{x_0+T}\left(2 \delta h\wedge \delta L-
2 L \delta h \wedge \delta h+\delta h_x\wedge \delta h\right) dx,
\eeq
where $\delta h=\delta g g^{-1}$. Note that the last term does not contribute
to the residues. The first two terms are meromorphic on $\G$ with poles
at $\g_s$ and $P_i\in D$, only. Therefore, a sum of their contributions
to residues of $\Omega'dz$ equals zero. Hence, $\omega$ does
descend to a form on
\beq\label{zz2}
\wh \P^D=\wh \P^D_0/\wh{GL_r}.
\eeq
Using (\ref{z19}) for the gauge transformation (\ref{agauge}), where
$\Phi_s$ depends on a point $z$ in the neighborhood of $\g_s$, we
obtain
\beq\label{z20}
\res_{\g_s} {\Omega} dz=-2\int_{x_0}^{x_0+T}
\left(\delta \kappa_s(x)\wedge\delta z_s(x)+\sum_{i=1}^r
\delta \b_s^i(x) \wedge\delta \a_s^i(x)\right)dx.
\eeq
From (\ref{z5}) we obtain that if $dz(P_i)=0$, then
\beq\label{zz3}
\res_{P_i}\Omega dz={\rm Tr}\left(\int_{x_0}^{x_0+T}
\left(\delta u^{(i)}(x)\wedge
\int_{x_0}^x \delta u^{(i)})(y)dy\right)dx-\delta u^{(i)}(x_0)\wedge
\int_{x_0}^{x_0+T}\delta u^{(i)}(x)dx\right).
\eeq
Equations (\ref{zz1}) imply that the restriction of $\delta u^{(i)}$ to
$\wh \P^D_0$ is $x$-independent. Then, from (\ref{zz3}) it follows that
the points $P_i\in \K$ give zero contribution to $\omega$.
From (\ref{z5}) and (\ref{z14}) it follows that the form
$\delta \Psi \Psi^{-1}$ when restricted to $\wh \P^D_0$
is holomorphic in the neighborhood of $P_m\in D_{\K}$. Therefore,
in this neighborhood
\beq\label{z21}
\left(\delta \Psi_x \Psi^{-1}+\delta \Psi \Psi_x^{-1}\right)
\Big|_{\wh \P^D_0}=0(1).
\eeq
Using this equality we obtain that on $\P_0^D$ the following
equation holds
\beq\label{z22}
\res_{P_m}\Omega dz=-2\res_{P_m}{\rm Tr}\left(\int_{x_0}^{x_0+T}
\left(L\delta \Psi\Psi^{-1}\wedge \delta \Psi\Psi^{-1}dx
\right)\right)dz.
\eeq
Therefore, restricted to $\wh \P_0^D$ the form $\omega$ is equal to the
integral over the period of (\ref{B3}).
The proof of the equation (\ref{z23}),
where $H_{a}$ is given by (\ref{z181}) is almost identical to the proof
of (\ref{H11}).

\noindent{\it Important remark.}
The formulae (\ref{z20}) and (\ref{z22}) do not directly imply that
$\omega$ restricted to $\wh \P$ is non-degenerate, because of the
constraints (\ref{zz1}). The conventional theory of the soliton equations,
and results of the next section provide some evidence that
it is non-degenerate for $D_{\K}\geq 0$,
although at this moment the author does not know a direct proof of that.
Anyway, equation (\ref{z23}) show that
the equations (\ref{z13}) are Hamiltonian on suitable subspaces of $\P^D$.
Then, commutativity of flows implies
\beq\label{z182}
\{H_a,H_b\}=0.
\eeq
The previous results can be easily extended for the case, when the leading
coefficient of the singular part of $L$ at the puncture $P_i$ has
multiple eigenvalues.
\begin{lem}
Let $L(x,w)$ be a formal Laurent series (\ref{z2})
such that $l_{-1}=C(x)u C^{-1}(x)$, and $u=u_i\delta^{ij}$ is a
constant diagonal matrix. Then there is a unique formal solution
$\Psi_0=\Psi_0(x,w)$ of the equation (\ref{z4}), which has the form
\beq\label{z5m}
\Psi_0(x,w)=C(x)\left(\sum_{s=0}^{\infty}\xi_s(x)w^{-s}\right) \E(x,w),
\ \ \E(x_0,w)=1,
\eeq
where
\beq\label{z6m}
\xi_{0}^{ij}=\delta^{ij};\ \ \xi^{ij}_s(x)=0, \ {\rm if} \ \
u_i=u_j,\ s\geq 1.
\eeq
and the logarithmic derivative $h(x,w)$ of $\E$ is a formal series
with non vanishing entries only for indices $(i,j)$, such that $u_i=u_j$,
i.e.
\beq\label{z51m}
h=\p_x\E\E^{-1}=
uw^{-1}+\sum_{s=0}^{\infty} h_s(x)w^s ,
\ \ h^{ij}=0, \ {\rm if}\ u_i\neq u_j
\eeq
The coefficients $\xi_s(x)$ of (\ref{z5m}) and the coefficients $h_s(x)$
of (\ref{z51m}) are differential polynomials of the matrix elements of $L$.
\end{lem}
Substitution of (\ref{z5m}) in (\ref{z4}) gives a system of
the equations which have the form (\ref{z9})
They recursively determines $\xi_{s+1}^{ij}$ for indices $(i,j)$ such that
$u_i\neq u_j$ and the matrix $h_s$, as polynomial functions of the
matrix elements of $l_i(x), \ i\leq s.$
\begin{cor} Let $\Psi_0$ be the formal solution (\ref{z5m}) of
equation (\ref{z4}). Then for any diagonal matrix $E=E_i\delta ^{ij}$
such that $E_i=E_j$, if $u_i=u_j$, the expression
$w^{-m}\Psi_0E\Psi_0^{-1}$ does not depend on $x_0$,
and is formally meromorphic.
The coefficients $m_s(x)$ of its Laurent expansion (\ref{z90})
are differential polynomials of the entries of the coefficients $l_i(x)$.
\end{cor}
The expression $w^{-m}\Psi_0E\Psi_0^{-1}$ is meromorphic and does not depend on
$x_0$ because $[\E,E]=0$.

The corollary, implies that if singular parts
of $L$ at the punctures $P_i$ have multiple eigenvalues, then the commuting
flows are parameterized by sets
\beq\label{f1m}
a=\left(P_i,m;E_{\l}\right),
\eeq
where $E_{\l}$ is a diagonal matrix that satisfies the condition of Corollary
5.2. The Hamiltonians of the corresponding equations are equal to
\beq\label{z183}
H_a=-\res_{P_i} {\rm Tr} \ \left(w^{-m}E_{\l}\ \int_0^T h(x)dx\right) dz.
\eeq

\medskip
\noindent{\bf Example. Field analog of the elliptic CM system.}

\medskip
\noindent Let us consider the zero curvature equation on the elliptic curve
with one puncture. We use the same notation as in section 4.
In the gauge $\a_s=e_s,\ e_s^j=\delta_s^j$,
the phase space can be identified with the space of  elliptic
matrix functions such that $L^{ij}$  has pole at the point $q_j(x)$ and
$z=0$, only. From (\ref{t21}) it follows that the residue of
$L^{jj}$ at $q_j$ equals $-q_{jx}$. Therefore,
$L^{jj}=p_j+q_{jx}(\zeta(z)-\zeta(z-q_j)-\zeta(q_j)$.
Equation (\ref{t21}) implies also, that $L^{ji}(q_j)=0,\ i\neq j$.

Let us assume, as in the case of the elliptic CM system,
that the singular part of $L$ at the puncture $z=0$ is a point of the orbit
of the adjoint action corresponding to the diagonal
matrix ${\rm diag}(r-1,-1,\ldots,-1)$. Then, taking into account
the momentum map corresponding to the gauge tarnsformation by diagonal
matrices, we get that the non-stationary analog of the Lax matrix for
the CM system has the form
\begin{eqnarray}
L^{ii}&=&p_i+q_{ix}\left(\zeta(z)-\zeta(z-q_i)-\zeta(q_i)\right),
\label{Wna}\\
L^{ij}&=&f_if_j{\s(z+q_i-q_j)\, \s(z-q_i) \s(q_j) \over
\s(z)\s(z-q_j)\,\s(q_i-q_j)\,\s(q_i)}, \ i\neq j.
\label{Wnb}
\end{eqnarray}
The values $f_i^2$ are fixed to
\beq\label{x0}
f_i^2=1+q_{ix},\ \ \ \sum_{i=1}^r q_{ix}=0.
\eeq
According to (\ref{z20}), the symplectic form
equals
\beq\label{x1}
\omega=\int_0^T\left(\sum_{i=1}^r \delta p_i(x)\wedge \delta q_i(x)\right)dx
\ \ \longmapsto \{p_i(x),q_j(y)\}=\delta_{ij}\delta(x-y).
\eeq
The commuting Hamiltonians are coefficients of the Laurent
expansion at $z=0$ of the quasimomentum, corresponding to the only simple
eigenvalue of the singular part of $L$ at $z=0$. To find them
we look for the solution of (\ref{z4}) in the form
\beq\label{x2}
\psi=C(x,z)e^{\int_0^x h(x',z)dx'},
\eeq
\beq
C=\left(\sum_{s=0}^{\infty} C^{(s)}(x)z^s\right),\ \
h=\sum_{s=-1}^{\infty}
h_s(x)z^s,
\eeq
where $C^{(0)}$ is the eigenvector of the singular part of $L$, corresponding
to the eigenvalue $(r-1)$, i.e.
\beq\label{x3}
C^{(0)}_i=f_i,
\eeq
and the coefficients $C^{(s)}$ for $s>0$ are vectors, normalized by the condition
\beq\label{x4}
\sum_{i=1}^rf_iC^{(s)}_i=0,\ \ s>0.
\eeq
Substitution of (\ref{Wna},\ref{Wnb}) into (\ref{z4}) gives a system of
the equations  for the coordinates $C_i$ of the vector $C$:
\begin{eqnarray}
\p_x C_i+hC_i&=&q_{ix}C_i\left[\zeta(z)-\zeta(z-q_i)-\zeta(q_i)\right]+
\nonumber\\
&{}&f_i\sum_{j\neq i}f_jC_j\left[\zeta(z)-\zeta(z-q_j)+\zeta(q_i-q_j)-
\zeta(q_i)\right],
\label{x5}
\end{eqnarray}
where we use the identity
\beq\label{x6}
{\s(z+q_i-q_j)\, \s(z-q_i) \s(q_j) \over
\s(z)\s(z-q_j)\,\s(q_i-q_j)\,\s(q_i)}=
\zeta(z)-\zeta(z-q_j)+\zeta(q_i-q_j)-\zeta(q_i).
\eeq
Taking the expansion of (\ref{x5}) at $z=0$, we find recursively the
coefficients of $C_i^{(s)}$ and densities $h_s$ of the Hamiltonians.
The first two steps are as follows.

The coeffcients at $z^{-1}$ of the right and left hand sides of (\ref{x5})
give
\beq\label{x7}
h_{-1}=q_{ix}+\sum_{j\neq i}f_j^2=
q_{ix}+(r-f_i^2)=r-1.
\eeq
The next system of equations is
\beq
f_{ix}+f_ih_0+(r-1)C_i^{(1)}=p_if_i+q_{ix}C_i^{(1)}+
f_i\sum_{j\neq i}\left(f_jC_j^{(1)}+f_j^2V_{ij}\right),
\eeq
where
\beq\label{x8}
V_{ij}=\zeta(q_j)+\zeta(q_{ij})-\zeta(q_i),\ \ q_{ij}=q_i-q_j.
\eeq
Using (\ref{x4}), we get
\beq\label{x9}
rC_i^{(1)}+f_ih_0=p_if_i-f_{ix}+f_i\sum_{j\neq i}f_j^2V_{ij}
\eeq
Multiplying (\ref{x9}) by $f_i$ and taking a sum over $i$, we find upon
using (\ref{x4}) and skew-symmetry of $V_{ij}$,
\beq\label{x10}
rh_0=\sum_{i=1}^r p_if_i^2=\sum_{i=1}^r p_i(1+q_{ix}).
\eeq
In the same way we get the system of equations for $C_i^{(2)}$
\beq\label{x11}
rC_i^{(2)}+\p_xC_i^{(1)}+h_0C_i^{(1)}+h_1f_i=
p_iC_i^{(1)}+q_{ix}f_i\wp(q_i)+
f_i\sum_{j\neq i}f_j\left(C_j^{(1)}V_{ij}+f_j\wp(q_j)\right).
\eeq
Consequently the expression for the density of the second Hamiltonian is
\begin{eqnarray}
r^2h_1&=&r\sum_{i}\left[q_{ix}f_i^2\wp(q_i)+
C_i^{(1)}\left(f_{ix}+p_if_i\right)
+\sum_{j\neq i}\left(f_i^2f_jC_j^{(1)}V_{ij}+f_i^2f_j^2\wp(q_j)\right)\right]
\nonumber\\
&=&r\sum_{i}\left[(r-1)f_i^2\wp(q_i)+C_i^{(1)}\left(f_{ix}+p_if_i+
\sum_{j\neq i}f_if_j^2V_{ji}\right)\right].\label{x13}
\end{eqnarray}
For the first line we have used the equation
$\sum_i \left(f_iC_{ix}^{(1)}+f_{ix}C_{ix}^{(1)}\right)=0$.
From (\ref{x9}) it follows that the second term in (\ref{x13})
equals
\beq\label{x14}
II=-rh_0^2+\sum_i\left(p_i^2f_i^2-f_{ix}^2+
\sum_{j\neq i} (f_i^2)_xf_j^2V_{ij}+\sum_{j,k\neq i} f_i^2f_j^2f_k^2
V_{ij}V_{ki}\right).
\eeq
For any triple of distinct integers $i\neq j\neq k\neq i$ the following
equation holds
\beq\label{x15}
V_{ij}V_{ki}+V_{jk}V_{ij}+V_{ki}V_{jk}=-\wp(q_i)-\wp(q_j)-\wp(q_k).
\eeq
In order to prove (\ref{x15}), it is enough to check that the left
hand side, which is symmetric function of all the variables $q_i,q_j,q_k$,
as a function of the variable $q_i$, has double pole at $q_i=0$,
and is regular at $q_i=q_j$.
In the same way one can obtain the well-known relation
\beq\label{x16}
V_{ij}V_{ji}=-\wp(q_i)-\wp(q_j)-\wp(q_{ij}).
\eeq
Equations (\ref{x15}, \ref{x16}) imply
\beq\label{x17}
\sum_i\sum_{j,k\neq i} f_i^2f_j^2f_k^2V_{ij}V_{ki}=-\sum_{i}\left(rf_i^2(r-f_i^2)
\wp(q_i)+\sum_{j\neq i}f_i^2f_j^4\wp(q_{ij})\right).
\eeq
From (\ref{x8}) it follows
\begin{eqnarray}
\sum_{j\neq i} (f_i^2)_xf_j^2V_{ij}=
{1\over 2}\sum_{j\neq i}\left((f_i^2)_xf_j^2-f_i^2(f_j^2)_x\right)
\left(\zeta(q_{ij})-2\zeta(q_i)\right)=\nonumber\\
=-{1\over 2}\sum_i \left(2r q_{ixx}\zeta(q_i)-\sum_{j\neq i}
q_{ijxx}\zeta(q_{ij})\right)+
{1\over 2}\sum_{j\neq i}\left(q_{ixx}q_{jx}-
q_{jxx}q_{ix}\right)\zeta(q_{ij}).  \label{x18}
\end{eqnarray}
The first sum is equal to
\begin{eqnarray}
{1\over 2}\sum_i \left(2r q_{ixx}\zeta(q_i)-\sum_{j\neq i}
q_{ijxx}\zeta(q_{ij})\right)=
{1\over 2}\sum_{i}\left(2r q_{ix}f_i^2\wp(q_i)-\sum_{j\neq i}q_{ijx}^2\wp(q_{ij})
\right)+\p_x F, \label{x19}
\end{eqnarray}
where
\beq\label{x20}
F={1\over 2}\sum_{i}\left(2rf_i^2\zeta(q_i)-
\sum_{j\neq i} q_{ijx}\zeta(q_{ij})\right).
\eeq
The first terms in (\ref{x13},\ref{x17},\ref{x18}) cancel each other.
The function $F$, as a function of the variable $q_i$, has poles at the points
$0, q_j, \ j\neq i$ and the sum of its residues at these points equals
\beq\label{x22}
rf_i^2-\sum_{j\neq i}q_{ijx}=r.
\eeq
Therefore, it has the same monodromy properties with respect to
all the variables. The functions $q_i(x)$ represent loops on the
elliptic curve. Therefore, $q_i(x+T)=q_i(x)+b_i$, where
$b_i$ is a period of the elliptic curve. The constraint (\ref{x0}) implies
$\sum_i b_i=0$. Then, from (\ref{x22}) it follows that $F$ is a {\it periodic}
function of $x$. The densities of the Hamiltonians are defined up to a total
derivative of periodic functions in $x$. Hence, a density of the second
Hamiltonian of the hierachy equals
\begin{eqnarray}\label{x23}
r^2\wt h_1=&-&{1\over r}\left(\sum_i(p_i(1+q_{ix})\right)^2+
\sum_i \left(p_i^2(1+q_{ix})-{q_{ixx}^2\over 4(1+q_{ix})}\right)\\
&-&{1\over 2}\sum_{j\neq i}\left((q_{ixx}q_{jx}-
q_{jxx}q_{ix}\right)\zeta(q_{ij})\\
&+&{1\over 2}\sum_{j\neq i}\left((1+q_{ix})(1+q_{jx})^2+
(1+q_{jx})(1+q_{ix})^2-q_{ijx}^2\right)\wp(q_{ij}).
\end{eqnarray}
The transformation $p_i\to p_i+f(x)$ does not change $h_s$ for $s>0$.
In particular, the first two terms in (\ref{x23}) can be rewritten as
\beq\label{x24}
-{1\over r}\left(\sum_i(p_i(1+q_{ix})\right)^2+
\sum_i p_i^2(1+q_{ix})={1\over 2r}\sum_{i,j} (p_i-p_j)^2(1+q_{ix})
(1+q_{jx}).
\eeq
The symplectic form (\ref{x1}) restricted to the subspace
\beq\label{x25}
\sum_iq_i=0, \ \ \sum_ip_i=0 ,
\eeq
is non-degenerate. The Hamiltonians $H_s=\int_0^T h_{s}(x)dx$ restricted
to this space generate a hierarchy of commuting flows, which we regard
as field analog of the elliptic CM system. For $r=2$  the Hamiltonian
$2H_1$ has the form (\ref{Ham}), where $q=q_1=-q_2, \ p=p_1=-p_2$.

\section{The algebro-geometric solutions}

So far, our consideration of the Bloch solutions (\ref{z16})
has been purely local and formal. For generic $L\in \A_0^D$ the series
(\ref{z5}, \ref{z51}) for the formal solutions $\Psi(x,q)$, and
quasimomentum  have zero radius of convergence. The main goal of this section
is to construct algebro-geometric solutions of the zero curvature equations,
for which these series do converge and, moreover, have meromorphic
continuations on a compact Riemann surface.

Let $\wh T(q)$ be a restriction of the monodromy operator $f(x)\to f(x+T)$
to the space of solutions of the equation $(\p_x-L(x,q)f=0$, where $f$
is a vector function. Then, we define  the Riemann surface $\wh \G$
of the Bloch solutions by the characteristic equation
\beq\label{R1}
R(\mu,q)\equiv\det\left(\mu-\wh T(q)\right)=
\mu^r+\sum_{j=1}^{r}R_j(q)\mu^{r-j}=0.
\eeq
\begin{lem} The coefficients $R_j(q)$ of the characteristic equation (\ref{R1})
are holomorphic functions on $\G$ except at the points $P_i$ of the divisor
$D$.
\end{lem}
{\it Proof.}
In the basis defined by columns of the fundamental matrix of solutions
to the equation $(\p_x-L)F(x,q;x_0)=0,\ \ F(x_0,q,x_0)=1$,
the operator $\wh T(q)$ can be identified with the matrix
\beq\label{R2}
\wh T(q)=F(x_0+T,q;x_0).
\eeq
{\it A'priory} this matrix is holomorphic on $\G$ except at the points of the divisor
$D$ and at points of the loops $\g_s(x)$, where $L$ has singularities.
From  Lemma 5.2 it follows that in the neighborhood of the loop we have
\beq\label{R3}
F=\Phi_s(x,q)\wt F(x,q)\Phi_s^{-1}(x,q),\ \wt F(x_0,q)=1,
\eeq
where $\wt F$ is a holomorphic matrix function, and $\Phi_s$ is
defined by (\ref{G1},\ref{G},\ref{G100}). The function $\Phi_s$ is periodic,
because $\g_s,\a_s$ are periodic.  In the neighborhood of the loop $\g_s$,
the functions $R_j(q)$ coincide with the coefficients of the characteristic
equation for $\wt F$. Therefore, they are holomorphic in that
neighborhood. The Lemma is then proved.

It is standard in the conventional spectral theory of periodic
linear operators that for a generic operator the Riemann surface
of the Bloch functions is smooth and has infinite genus. For
algebro-geometric or finite-gap operators the corresponding Riemann surface
is singular, and is birational equivalent to a {\it smooth} algebraic curve.

It is instructive to consider first, as an example of such operators,
the case, when $L$ does not depend on $x$, i.e. $L\in \L^D$.
In this case the equation $(\p_x-L)\psi=0$ can be easily solved.
The Bloch solutions have the form
\beq\label{R4}
\psi=\psi_0 e^{kx},
\eeq
where $\psi_0$  is an eigenvector of $L$, and $k$ is the correspoding
eigenvalue. These solutions are parameterized by points $Q$ of the spectral
curve $\wh \G_0$ of $L$. The image of $\wh \G_0$ under the map into
${\bf C}^1\times \G$ defined by formula
\beq\label{R5}
(k,q)\in \wh \G_0 \longmapsto \  (\mu=e^{kT},q)\in {\bf C}^1\times \G
\eeq
is the Riemann surface $\wh \G$ defined by (\ref{R1}), where the coefficients
are symmetric polynomials of $e^{k_i(q)T}$.

For example, if $\wh \G_0$ is defined by the equation
\beq\label{R51}
k^2+u(q)=0,
\eeq
where $u(q)$ is a meromorphic function with double poles at the points of $D$,
then $\wh \G$ is defined by the equation
\beq\label{R6}
\mu^2+2R_1\mu+1=0, \ \ \ R_1(q)=\cosh \ (\sqrt {u(q)}).
\eeq
The Riemann surface defined by (\ref{R6}) is singular. Projections
onto $\G$ of the points of self-intersection of $\wh \G$ are roots of the
equation
\beq\label{R7}
u(q)=\left({\pi N\over 2T}\right)^2,
\eeq
where $N$ is an integer. The coefficient $u(q)$ has poles of the second
order at $D$, $u=a_i^2w^{-2}+O(w^{-1})$, where $w$ is a local coordinate
at $P_i\in D$. Therefore, as $|N|\to \infty$, the roots of (\ref{R6}) tend
to the points of $D$. The coordinates of the singular points $q_{i,N}$,
that tend to $P_i$ equal
\beq\label{R8}
w (q_{i,N})=2Ta_i (\pi N)^{-1} + O(N^{-2}).
\eeq
As usual in perturbation theory, for generic $L$
each double eigenvalue $q_{i,n}$ splits into two smooth branch points
$q_{i,n}^{\pm}$. By analogy with the conventional theory we expect, that if
$L$ is an analytic  function of $x$, then the differences
$|w (q_{i,N})-w (q_{i,N}^{\pm})|<O(N^{-k})$ will decay faster that any power
of $N^{-1}$.

Localization of the branch points is a key element in the construction
\cite{mckean} of a theory of theta-functions for infinite genus hyperelliptic
curves of the Bloch solutions for periodic Sturm-Liouville operators.
In \cite{kr8} a general approach for the construction
of Riemann surfaces of the Bloch functions was proposed. The model of
the spectral curves developed in \cite{kr8} was chosen in \cite{trub}
as a starting point of the theory of general (non-hyperelliptic)
infinite-genus Riemann surfaces. It was shown that for such surfaces
many classical theorems of algebraic geometry take place.

Algebro-geometric or finite-gap operators can be seen, as operators
for which there are only a finite number of multiple eigenvalues that
split into smooth branch points. Let $\wh \G$ be a smooth genus $\wh g$ algebraic curve
that is an $r$-branch cover of $\G$. Note that, unlike the stationary case,
for given a rank $r$ their is no relation
between $\wh g$, and the genus $g$ of $\G$. As $\wh g$ increases
the dimension of the  space of $r$-sheeted cover increases. It equals
$2(\wh g-rg+r-1)$.

Assume that the preimages $P_i^l, P_0^l$ on $\wh \G$
of the points of a divisor $D$, and a point $P_0$ on $\G$ are not
branch points. The definition of the Baker-Akhiezer function corresponding
to this data and to a non-special degree $\wh g+r-1$ divisor
$\wh \g$ on $\wh \G$ is as follows:

$1^0$. $\psi$ is a meromorphic vector function on $\wh \G$
except at the points $P_i^l$. Its divisor of poles on $\wh \G$ outside of
$P_i^l$ is not greater that $\wh \g$;

$2^0$. in the neighborhood of $P_i^l$ the vector function $\psi$ has the form
\beq\label{y1}
\psi=\xi_{i,l}(q,t)
\exp\left(\sum_m t_{(i,m;l)}w^{-m}\right),
\eeq
where $\xi_{i,l}(q,t)$ is a holomorphic vector-function;

$3^0$. evaluation of $\psi$ at the punctures $P_0^l$ are vectors with
coordinates $(\psi(P_0^l))^{(i)}=\delta^{il}.$
\begin{th}
Let $\psi(q,t)$ be the Baker-Alhiezer vector function associated with
a non-special divisor  $\wt \g$ on $\wh \G$. Then, there exist unique
matrix functions $M_{(i,m;l)}(q,t)\in \N^D_{\g(t),\a(t)}$ such that the
equations
\beq\label{y2}
\left(\p_{(i,m;l)}-M_{(i,m;l)}\right)\psi(q,t)=0
\eeq
hold.
\end{th}
Now, let $v^{(i)}_l(x)$ be a set of periodic functions,
$\int_{0}^Tv^{(i)}_ldx=0$, and $u_l^{(i)}$ be a set of constants.
Then the change of the independent variables
\beq\label{y33}
t_{(i,1;l)}=xu^{(i)}_l+v^{(i)}_l(x)+t'_{(i,1;l)}
\eeq
define the Baker-Akhiezer function $\psi$, as a function of $(q,t)$ and the
variable $x$. From (\ref{y2}) it follows that
\beq\label{y3a}
(\p_x-L)\psi=0,\ \ \ L=\sum_{i,l}\left(u^{(i)}_l+\p_xv^{(i)}_l\right) M_{(i,1;l)}
\eeq
As follows from Lemma 5.2,
the vector $\D(M_a), \ a=(i,m;l)$, corresponding to $M_{a}$
under (\ref{map}), is tangent $(\g_s(t_{a}),\a_s(t_{a}))$.
Therefore, $\D(L)$ is tangent to $(\g_s(x),\a_s(a))$.

In general, $L$ constructed above is not a periodic function of $x$.
It is periodic, if we impose additional constraints on the set of
data that are the curve $\wh \G$ and the constants $u^{(i)}_l$.
We call the set $\{\wh \G,u^{(i)}_l\}$ admissible if
there exists a meromorphic differential $dp$ on $\wh \G$
which has second order poles at $P_i^l$
\beq\label{y3b}
dp=-u^{(i)}_ldw\left(w^{-2}+O(1)\right),
\eeq
and such that all periods of $dp$ are multiples of $2\pi i/T$
\beq\label{y4}
\oint_{c} dp={2\pi i m_c\over T},\ \ m_c\in Z, \ \ c\in H_1(\G,Z).
\eeq
\begin{lem} The Baker-Akhiezer function $\psi$, associated with
an admissible set of data $\{\wh \G,u^{(i)}_l\}$
satisfies the equation
\beq\label{y5}
\psi(x+T,q)=g\psi(x,q)\mu(q),\ \ \mu=e^{p(q)T},
\eeq
where $g$ is the diagonal matrix $g={\rm diag} \
(\mu(P_0^1),\ldots,\mu(P_0^r))$.
\end{lem}
From (\ref{y4}) it follows that the function $\mu$ defined by the multi-valued
abelian integral $p$ is single-valued. Equation (\ref{y5}) follows from
the uniqueness of the Baker-Akhiezer function, because the left and the
right hand sides have the same analytic properties.

The matrix function $L$ constructed with the help of $\psi$ satisfies
the monodromy property
\beq\label{y6}
L(x+T,q)=gL(x,q)g^{-1}.
\eeq
Let $\SP=\SP(T,p)$ be a space of curves $\wh \G$ with meromorphic differential
$dp$ satisfying (\ref{y4}). We would like to mention, that the closure of
$\SP$, as $T\to \infty$, coincides with the space of all genus $\wh g$
branching covers of $\G$.
\begin{cor} A set of data $\wh \G\in \SP, [\wh \g]\in J(\wh \G)$, and
a set of periodic functions $v_l^{(i)}(x)$ define with the help of the
corresponding Baker-Akhiezer function a solution of the hierarchy (\ref{z13})
on $\B^D/\wh{GL_r}$.
\end{cor}
The finite-gap or algebro-geometric solutions are singled out by
the constraint that there is a Lax matrix $L_1\in \L^{nD}$ such that
\beq\label{y71}
[\p_x-L, L_1]=0.
\eeq
Indeed, let $k$ be a function on $\wh \G$ with divisor of poles $n\wh D$,
where $\wh D$ is the preimage of $D$. If $n$ is
big enough this exists. Let $\psi$ be the Baker-Akhiezer function on
$\wh \G$, then as it was shown above there is a unique Lax matrix $L_1$
such that
\beq\label{y8}
L_1(t,q)\psi(t,q)=k(q)\psi(t,q).
\eeq
Equation (\ref{y8}) implies that the spectral curve of $L_1$ is birationally
equivalent to the Riemann surface $\wh \G$ of Bloch solutions for $L$.
\begin{th} The form $\omega$ defined
by (\ref{form}) and (\ref{z15}) restricted to the space of algebro-geometric
solutions, corresponding to a set of function $v_l^{(i)}(x)$
equals
\beq\label{a10}
\omega=\sum_{s=1}^{\wh g+r-1} \delta p(\wh \g_s)\wedge \delta z(\wh \g_s).
\eeq
\end{th}
The meaning of the right hand side of this formula is analogous to
that of formula (\ref{d1}). It shows that the form $\omega$ restricted
to the space of algebro-geometric solutions is non-degenerate.

It is well-known, that the finite-gap solutions of the KdV hierarchy
are dense in the space of all periodic solutions (\cite{mar}).
As shown in \cite{kr8} the finite-gap solutions are dense for the KP-2
equation as well. It seems quite natural to expect that the similar result
is valid for the zero-curvature equations on an arbitrary algebraic curve, as
well. In the conjectued scenario the infinite dimensional space $\B^D$ can be
identified with a direct limit of finite-dimensional spaces
$\L^{nD}$, as $n\to \infty$. We are going to address that problem in the
near future.

\end{document}